\documentclass[reprint,amsmath,amssymb,superscriptaddress,prc,floatfix]{revtex4-1}
\usepackage{amsmath}
\usepackage{graphicx}
\usepackage{longtable}
\usepackage{bm}
\usepackage[caption=false,subrefformat=parens]{subfig}
\usepackage{hyperref}
\hypersetup{colorlinks=true,breaklinks,linkcolor=red,citecolor=blue}

\begin{document}
  
\title{Fission dynamics within time-dependent Hartree-Fock: Boost-induced fission}

\author{Philip Goddard}
\affiliation{Department of Physics, Faculty of Engineering and Physical Sciences, University of Surrey, Guildford, Surrey GU2 7XH, United Kingdom}

\author{Paul Stevenson}
\email{p.stevenson@surrey.ac.uk}
\affiliation{Department of Physics, Faculty of Engineering and Physical Sciences, University of Surrey, Guildford, Surrey GU2 7XH, United Kingdom}

\author{Arnau Rios}
\email{a.rios@surrey.ac.uk}
\affiliation{Department of Physics, Faculty of Engineering and Physical Sciences, University of Surrey, Guildford, Surrey GU2 7XH, United Kingdom}

\date{\today}

\begin{abstract}
\begin{description} 
\item[Background] Nuclear fission is a complex large-amplitude collective decay mode in heavy nuclei. Microscopic density functional studies of fission have previously concentrated on adiabatic approaches based on constrained static calculations ignoring dynamical excitations of the fissioning nucleus, and the daughter products.
\item[Purpose] We explore the ability of dynamic mean-field methods to describe induced fission processes, using quadrupole boosts in the nuclide $^{240}$Pu as an example.
\item[Methods] Following upon the work presented in Ref.~\cite{Goddard2015}, quadrupole-constrained Hartree-Fock calculations are used to create a potential energy surface. An isomeric state and a state beyond the second barrier peak are excited by means of instantaneous as well as temporally extended gauge boosts with quadrupole shapes. The subsequent deexcitation is studied in a time-dependent Hartree-Fock simulation, with emphasis on fissioned final states. The corresponding fission fragment mass numbers are studied. 
\item[Results] In general, the energy deposited by the quadrupole boost is quickly absorbed by the nucleus. In instantaneous boosts, this leads to fast shape rearrangements and violent dynamics that can ultimately lead to fission. This is a qualitatively different process than the deformation-induced fission. Boosts induced within a finite time window excite the system in a relatively gentler way, and do induce fission but with a smaller energy deposition. 
\item[Conclusions] The fission products obtained using boost-induced fission in time-dependent Hartree-Fock are more asymmetric than the fragments obtained in deformation-induced fission, or the corresponding adiabatic approaches. 
\end{description}
\end{abstract}


\maketitle

\section{Introduction}

Induced fission processes are of particular practical relevance in a variety of environments, ranging from energy production to nuclear waste disposal and astrophysics \cite{Spe74,Wagemans1991,Byrne1994,gre96,Kra12,Hamilton2013}. Experimental data are necessary to access observables such as mass yields or excitation energies, of the fission process in a wide range of isotopes. Where no experimental data are available, however, a robust, predictive theory is still needed. In contrast to traditional models that require phenomenological input, theoretical studies based on microscopic inputs provide invaluable information on qualitatively relevant aspects of fission, e.g., new fission modes \cite{And13,Wak14}. 

The theoretical description of the induced fission process is often discussed within a Bohr and Wheeler framework \cite{Boh39}. At a first stage, one assumes that a nucleus is excited by the absorption of an incident neutron. The corresponding compound nucleus subsequently decays and breaks into two (or more) fragments that, in turn, decay to their respective ground states in a variety of ways. Phenomenological models based on these ideas have been successful at describing different aspects of fission phenomena, see Refs.~\cite{Mol01,Mol04,Randrup2011,Mol12,Rod14} for recent advances.  

Ideally, a microscopic description of the fission process should start by accounting the individual existence of neutrons and protons, as well as their interactions. Because of the large-amplitude nature of the fission process and the rapid timescales associated with it,  individual and collective dynamics need to be considered explicitly. Starting at this microscopic level and building the description of the fission process from the ground up in a time-dependent theory, one should, in principle, account for quantal and dissipative processes, fluctuations, correlations, and dynamical effects that semi-phenomenological models can only include in an \emph{ad hoc} manner. Nuclear density functional theory in its time-dependent formulation, which we hereafter call Time-Dependent Hartree-Fock (TDHF), is a good starting point for such a microscopic theory of fission phenomena \cite{Uma10,Goddard2015,Scamps2015,Bulgac2015}. 

On the one hand, advances in computational power allow for relatively straightforward calculations of ground states, including, if necessary, shape constraints of heavy and superheavy nuclei in fully unrestricted three-dimensional (3D) geometry \cite{Mar13,Pei2014,Ryssens2015}. This approach provides access to a potential energy surface (PES) that is dictated by the energy density functional (EDF) alone \cite{Egido2000,War02,McD13,McD14a,Rod14}. On the other hand, simulations of the unrestricted time evolution of nuclei are now possible with a variety of TDHF solvers \cite{Uma10,Goddard2015,Scamps2015,Bulgac2015}. Consequently, one can now take predictions of the PES as starting points in a dynamical evolution that mimics a fission process, taking into account both single-particle and collective dynamics. This represents a step forward from the traditional picture based on adiabatic approaches. 

The TDHF approach, however, presents a number of limitations, and the simulated fission processes should thus only be considered as proxies of the actual fissioning system. The correct time evolution of heavy nuclei should incorporate pairing correlations, and hence a superfluid description is desirable \cite{Ste11,Scamps2015,Tanimura2015}. Symmetry breaking and multi-configurational correlations lie beyond the scope of TDHF \cite{Gou05,Bernard2011}, but are relevant for the mass region of interest for fission. The final fragments obtained within such an approach do not have integer mass numbers, and projection into good particle number is needed if meaningful mass distributions are to be extracted from simulations \cite{Sim10,Scamps2015}. Perhaps more importantly, collective tunneling is not explicitly incorporated in the theory and one is therefore hampered in predicting realistic fission life-times \cite{Bertsch2015}.

In spite of these shortcomings, the combination of shape-restricted ground-state calculations together with dynamical TDHF simulations provides significant insight into the fission process. In a previous publication, we have used this approach to study the pre- and postscission dynamics of $^{240}$Pu \cite{Goddard2015}. In quadrupole-constrained calculations, we identified three regions of the PES with very different dynamics. First, below the second barrier maximum, fission is forbidden within TDHF time scales. Time-evolved states exhibit complex oscillatory dynamics in line with giant resonances \cite{Goddard14}. One can thus interpret the evolution in the forbidden region as rapid oscillatory motion around local minima in a generalized PES.

Second, as deformation increases shortly after the maximum in the PES, fission still does not occur within the time scale of a TDHF calculation. Dynamical simulations in this region show large-amplitude oscillations, with the nucleus slowly exploring a range of deformation parameters. We interpret these oscillations as pre-scission vibrations, driven by the Coulomb interaction between two lobes of the compound nucleus. In spite of their relative violent nature, these vibrations cannot fission the nucleus within time scales of $10^3-10^4$ fm/$c$, possibly owing to the lack of freedom in exploring possible pathways caused by the lack of pairing correlations \cite{Scamps2015}.

Third, we observe that fission is allowed in dynamical calculations beyond the crossing of the one- and the two-fragment pathways in the PES. Energetic arguments explain the appearance of this allowed region within TDHF. The deformation-induced fission (DIF) process in the allowed region can be interpreted as a surrogate of spontaneous fission. The initial states of the dynamics represent, to some extent, different post-tunneling configurations in the PES. While their dynamics should be influenced by pairing and correlation effects, the mere fact that different kinds of fragments are obtained dynamically as the outer section of the PES is explored highlights the importance of non-adiabatic effects in this process.

 We use the same numerical framework devised to treat the pre-fissioned state to analyze the properties of the outgoing fission fragments. DIF fragments are more asymmetric than the fission products predicted with the corresponding adiabatic, two-fragment pathway. The total excitation energy of the fission process can be determined by either extrapolating the total collective kinetic energy, or comparing the (excited) fragment energies to their ground-state counterparts. Moreover, slowing down the corresponding fragments using a Galilean boost, we study their internal excitations and determine their corresponding excitation spectra. All these technical developments, summarized in Ref.~\cite{Goddard2015} and explained in detail in Ref.~\cite{Goddard14}, allow for a detailed study of the fission process, from the prefission to the postscission phase, within a single coherent theoretical framework. 

This paper is concerned with investigating methods which induce fission for initial configurations where the process is either forbidden or inhibited within the time scale of TDHF. This is referred to as boost-induced fission (BIF), in contrast to the cases of DIF presented in our previous paper \cite{Goddard2015}. In the same spirit, we are not so much concerned with the detailed predictive power of our approach. Instead, we want to explore the potential of dynamical calculations in the context of induced fission reactions. To this end, we consider two different energy-deposition processes that will mimic an initial excitation of the pre-scission system. 

Large-amplitude collective motion may be induced by applying an external field to the system. A reasonable choice for this external field is one which will provide a quadrupole excitation, in line with Refs.~\cite{Oko83,Jun88}. The choice of a quadrupole shape is motivated by the knowledge that fission requires at least in part an increase in quadrupole deformation.
Hence, we apply a quadrupole gaugelike transformation to excite nonfissioning states into a fissioned configuration. First, we study external excitation fields which are applied instantaneously, and thus mimic quick energy depositions. We also analyze energy depositions that are simulated with a time-dependent profile. We are not particularly concerned with the physical mechanism causing the energy deposition. Naively, instantaneous quadrupole boosts are related to fast deformation-inducing processes, e.g., high-energy particle absorption. In contrast, time-modulated quadrupole boosts induce slow shape changes. In an over-simplistic picture that does not consider geometrical aspects, one could associate these with slow, thermal neutrons which are absorbed by a slowly rearranging nucleus. We note that excited initial states can also be useful in the framework of $\beta$-delayed fission processes, where one envisages an excited daughter nucleus as a starting point of the fission process \cite{And10,War12,And13}. 

This paper is organized as follows. In Sec.~\ref{sec:technical}, we give a brief review of the numerical implementation of the BIF approach. Section \ref{sec:instant} discusses the fission process generated by instantaneous quadrupole boosts, whereas Sec.~\ref{sec:ext} is focused on time-dependent energy deposition. We analyze the masses of the fission products in Sec.~\ref{sec:massdistbifdif}. Concluding remarks and an outline for future research are given in Sec.~\ref{sec:conclusions}.

\section{Technical implementation} 
\label{sec:technical}

As in our previous paper, we use the fission benchmark isotope $^{240}$Pu \cite{Bertsch2015}. Data for the spontaneous  \cite{Wat62,Tor71} and induced \cite{Jfis} fission of this isotope is available. The technical implementation of our ground-state and dynamical calculations is the same as in Ref.~\cite{Goddard2015}, with further details provided in Ref.~\cite{Goddard14}. Arbitrarily quadrupole shape-constrained ground-states have been obtained by means of an augmented Lagrangian method \cite{Sta10} with a suitable masking procedure. The SkM$^\ast$ effective interaction is used throughout \cite{Bar82}. The grid used in static calculations has $40^3$ points, ranging from $-19.5$ to $19.5$ fm in the $x,y,$ and $z$ directions. Time-dependent calculations were performed in a grid of $48\times48\times160$ points from $-23.5$ to $23.5$ fm in the $x$ and $y$ directions and from $-79.5$ to $79.5$ fm in the $z$ direction.

We include BCS pairing within the static calculation, using a Volume-$\delta$ interaction \cite{Mar13}. The corresponding ground-state properties are very close to other results in the literature. The TDHF dynamical evolution is simulated using the {\sc sky3d} code \cite{Mar13}. Pairing beyond a frozen-occupation approximation is not included. The lack of single-particle occupation rearrangement can lead to relatively artificial fission fragments. We note, however, that in this BIF study the pre-scission fragment is excited energetically and hence pairing effects are expected to play a smaller role than in the DIF case.

Dynamical nuclear observables are computed by applying a suitable comoving spatial mask that takes into account the two-fragment nature of the scissioned products. The periodic nature of our boundaries could potentially cause some artificial effects. Somewhat computationally expensive methods have been devised to treat the continuum problem within TDHF \cite{Par13,Par14,Bas15}, but they are not implemented here. In general, we have found that particle emission plays a very small role. The decay in total particle number is of the order of $0.1-0.2$ particles during the postscission evolution. As a matter of fact, postscission fragments have fluctuations in particle number below the $0.05$ level when our mask procedure is implemented. The nearest integer mass numbers thus obtained are thus free of boundary errors.

\begin{figure}[t!]
\begin{center} 
\includegraphics[width=\linewidth]{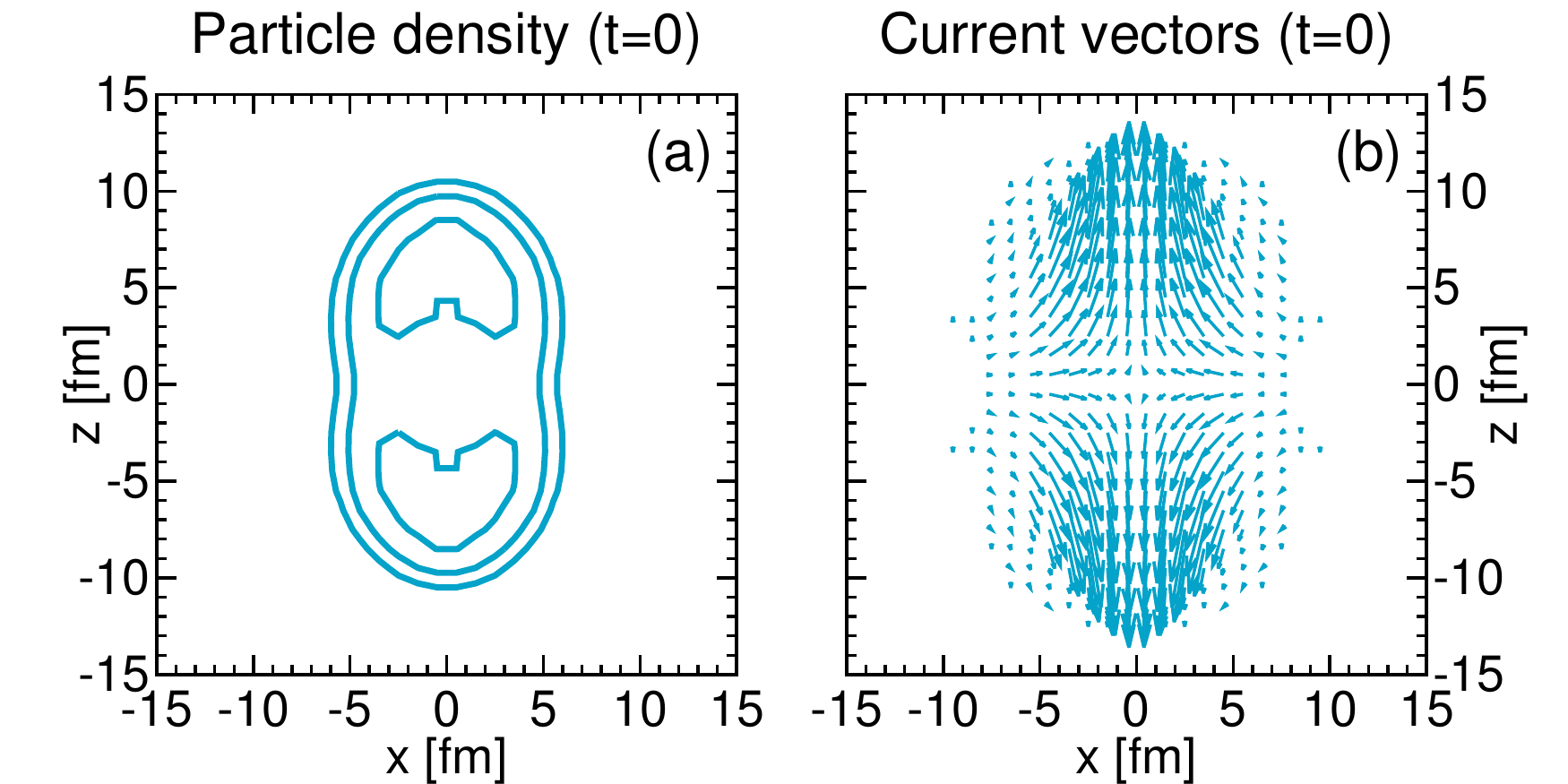} 
\caption{(Color online) (a) Slice of the 3D particle density. The isolines are separated by $0.05$ particles/fm$^3$. (b) Current vectors, $\bm j (\bm r)$, for a quadrupole velocity field applied instantaneously. Both pictures are taken for the isomeric state, $\beta_{20}=0.682$, at time $t=0$. The current vectors have been normalized to a visually instructive length.}
\label{is_quad_curr}
\end{center}
\end{figure} 

\section{Instantaneous boosts}
\label{sec:instant}

We simulate an excitation of the system by means of a gauge transformation, $e^{i\eta \phi(\bm r)}$, applied to the initial-state single-particle wave functions. This corresponds to a velocity boost which carries the profile $\vec v \sim \eta \nabla \phi(\bm r)$.

\begin{figure}[t!]
\begin{center} 
\includegraphics[width=0.6\linewidth]{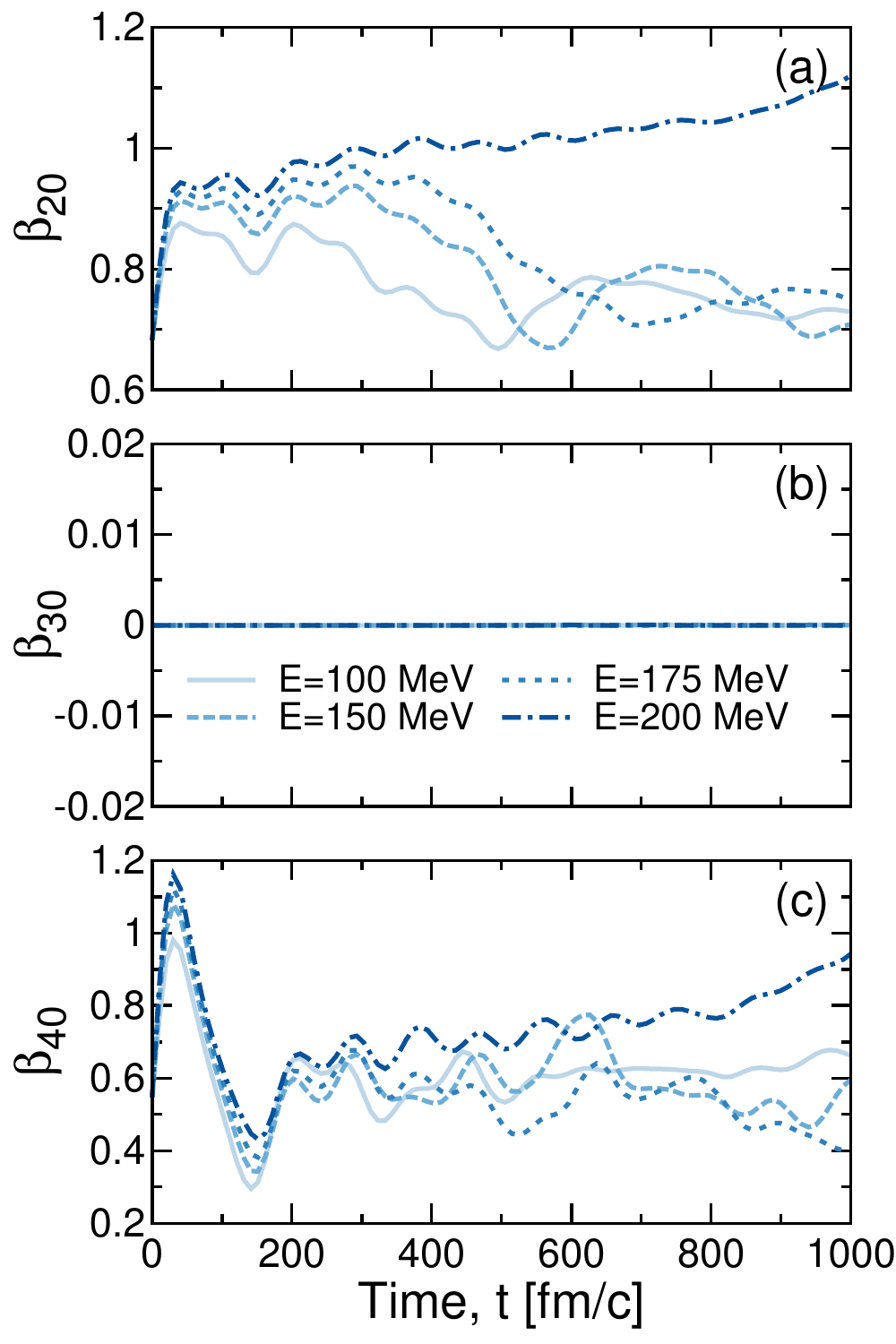} 
\caption{(Color online) Time evolution of (a) quadrupole, (b) octupole and (c) hexadecapole deformation parameters following instantaneous quadrupole excitations upon the isomeric state. The threshold for fission is between $175$ and $200$ MeV. Scission occurs between $950$ and $1000$ fm/$c$ for the $E=200$ MeV boost.}
\label{is_quad_boost}
\end{center}
\end{figure}

\begin{figure*}[t!]
\begin{center} 
\includegraphics[width=0.6\linewidth]{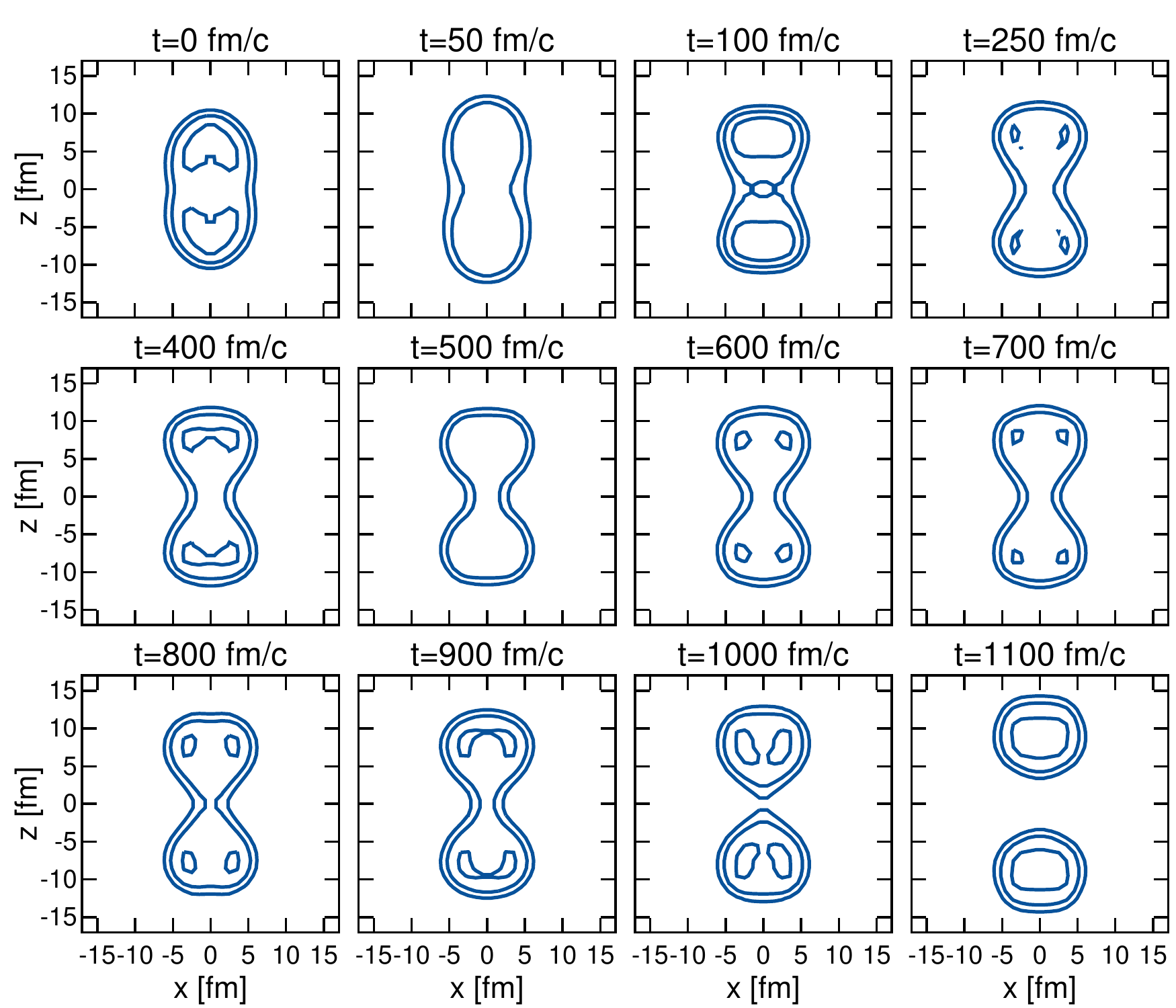} 
\caption{(Color online) Slices of the total particle density for various times, following an instantaneous $200$-MeV quadrupole excitation upon the isomeric state. The isolines are separated by $0.05$ particles/fm$^3$.}
\label{is_quad_boost_dens}
\end{center}
\end{figure*} 

The parameter $\eta$ determines the intensity of the boost. Owing to the gauge invariance of the Skyrme and Coulomb interactions, the boost deposits only collective kinetic energy \cite{Dob96,Dob05}. For instantaneous boosts, the exact amount of deposited kinetic energy is easily computed (see the Appendix). One can thus adjust $\eta$ to excite the nucleus with a given amount of collective energy.  The spatial profile, $\phi (\bm r)$, is chosen here to be proportional to a quadrupole field, as in some previous studies~\cite{Oko83,Jun88}. 

Two initial configurations will be investigated: the fission isomer of $^{240}$Pu at $\beta_{20}=0.682$, and a point just beyond the peak of the second fission barrier at $\beta_{20}=0.890$.

\subsection{Instantaneous BIF on the isomeric state}

Figure~\@ \ref{is_quad_curr} shows the initial density [panel (a)] and current [panel (b)] for the case of the quadrupole boost acting upon the fission isomer. The quadrupole boost produces a current that will initially pull away the two lobes of the particle density in opposite directions. The intensity of the boost will determine the strength of these currents and the corresponding amount of collective energy deposited by the boost. 

Figure~\ref{is_quad_boost} shows the evolution of the deformation parameters following a quadrupole velocity boost upon the isomeric state applying different amounts of energy in the region between $100$ and $200$ MeV. These relatively large energies are chosen to explore both configurations that remain bound or that fission within a time scale of $\approx \! 1000$ fm/$c$. In all cases, an initial, rapid increase in quadrupole deformation is found within the first $50$ fm/$c$, as shown in panel (a). As expected, the quadrupole rise is larger for a larger energy deposition. Following the quick increase in elongation, for an excitation below $175$ MeV, the nucleus draws back to its original quadrupole deformation, and then begins low-frequency, large-amplitude vibrations. As the initial configuration is mass symmetric, and the excitation was of a pure quadrupole nature, no octupole deformation is induced, as evinced in panel (b). 

Inspection of Fig.~\ref{is_quad_boost} leads to a threshold energy for instantaneous BIF.  An unbounded increase in the quadrupole moment indicates fission, and one observes that the threshold energy for an instantaneous quadrupole boost is $175 < E_{\text {thresh}} \le 200$ MeV. For $E=200$ MeV, in contrast to the lower energy cases, the quadrupole deformation gradually increases, while oscillating, as the system moves to a fissioned configuration. This differs from the evolution towards fission in the DIF cases studied in Ref.~\cite{Goddard2015}. There, the system smoothly evolved to a fissioned configuration by increasing steadily $\beta_{20}$, with no oscillations. The oscillations on the quadrupole degrees of freedom may be interpreted as a consequence of the rapid, large energy deposition. We comment upon these shortly.

In all cases, the evolution of the hexadecapole deformation of Fig. \ref{is_quad_boost}(c) demonstrates that the nucleus necks significantly between $100$ and $150$ fm/$c$. This is reflected in a characteristic drop in magnitude of $\beta_{40}$ as the elongation increases. We note that this trait is also found in the evolution with quadrupole moment of constrained Hartree-Fock (CHF) calculations \cite{Rod14,Goddard2015}. In contrast, fission induced dynamically by deformation did not show a stark decrease of $\beta_{40}$, as the initial configurations were deformed such that they were already displaying significant necking \cite{Goddard2015}.

Figure \ref{is_quad_boost_dens} displays snapshots of slices of the 3D particle density after the excitation of $200$ MeV is applied. Following a sharp increase in quadrupole shape until about $50$ fm/$c$, the nucleus oscillates with a period of about $\approx 100$ fm/$c$. Two symmetric pre-formed fragments can be distinguished early on. The compound nucleus oscillates in shape, while the quadrupole moment increases, while oscillating. Around $900$ fm/$c$, the Coulomb repulsion between the two lobes is strong enough to bring the configuration to scission. The resulting two fragments are equal in mass. 

These results contradict the naive assumption that the instantaneous quadrupole excitation will simply move the nucleus by the corresponding energy along the static PES. The static PES fission barriers are of the order of $\approx10$ MeV. In contrast, the energy required for BIF is an order of magnitude larger. One can attribute this to the gauge-invariant nature of our choice of boost. All of the energy is imparted in the form of collective kinetic energy so that both the static and dynamic cases have the same potential energy at $t=0$. The instantaneous boost causes the initial state to depart from the static PES at $t=0$, as the dynamic state now contains considerable internal excitation. In a sense, we are exploring the PES in a new dimension by modifying the kinetic content of the EDF.

\begin{figure}[h!]
\begin{center} 
\includegraphics[width=\linewidth]{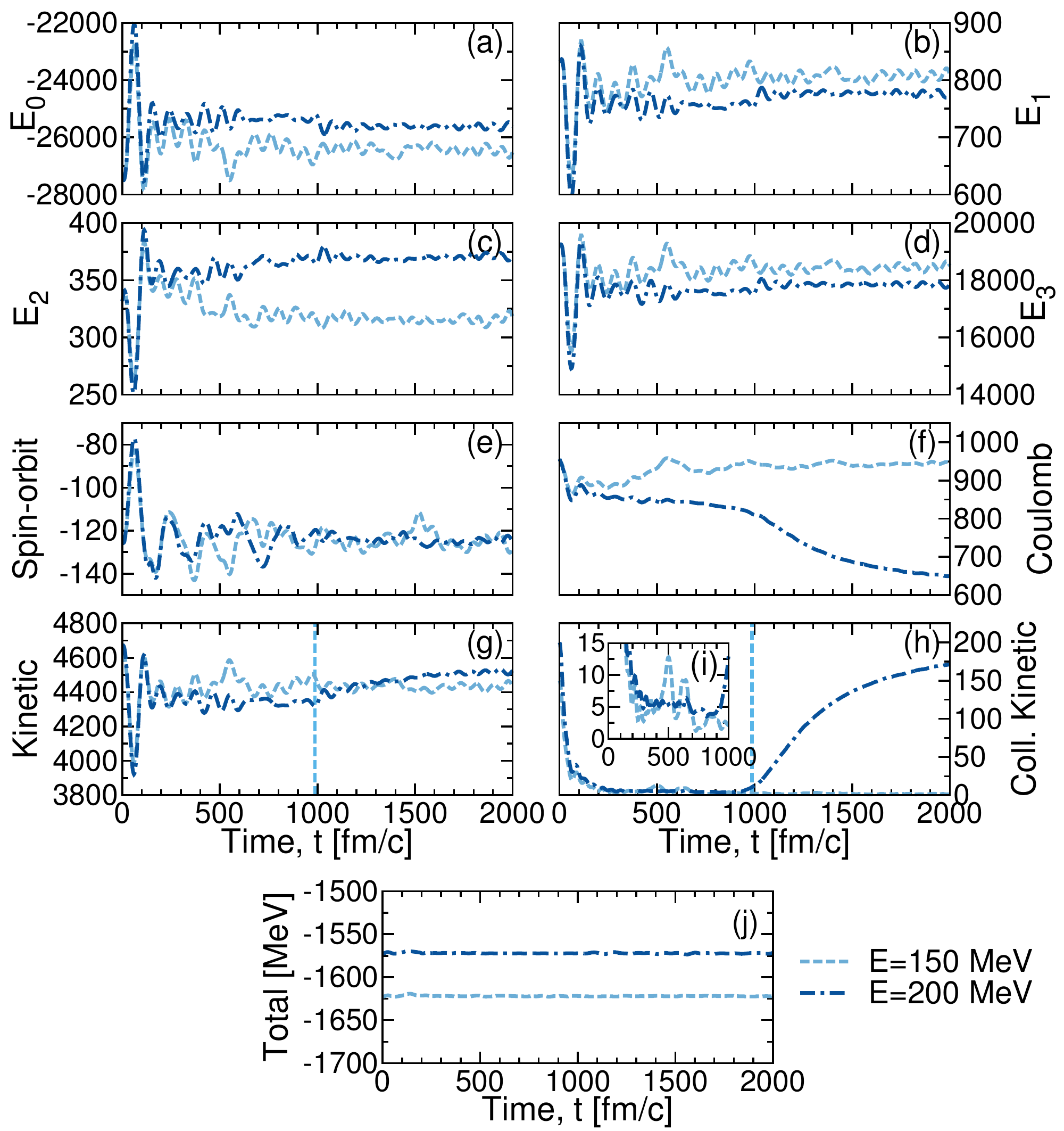} 
\caption{(Color online) (a)-(j): Evolution of the integrated contributions to the EDF following instantaneous quadrupole excitations upon the isomeric state. Vertical lines in panels (g) and (h) correspond to the point of scission. Units in all panels are MeV. See text for more details.} 
\label{symm-iso-fiss}
\end{center}
\end{figure} 

One can gain further insight into the BIF mechanism by analyzing the time evolution of the different contributions to the EDF. The individual terms of the functional can be used to pinpoint physical processes. We show a series of EDF terms in Fig.\@ \ref{symm-iso-fiss}, comparing a nonfissioning (dashed line) to a fissioning (dot-dashed line) configuration \cite{Goddard2015}. The terms of the functional are defined in Ref.~\cite{Mar13}. For both cases, other than the collective and total energy, all of the contributions to the energy functional are initially identical. 

For both configurations, the initial boost energy is imparted to the system as pure collective kinetic energy at $t=0$, as shown in Fig.~\ref{symm-iso-fiss}(h). The energy  provided by the excitation field is rapidly transferred into the nuclear terms of the energy functional. By $200$-$250$ fm/$c$, a roughly constant collective energy of $\approx5$ MeV remains for both configurations [inset panel (i)].
This energy corresponds to the relatively disruptive internal currents induced by the boost. For the fissioning case, the collective energy ramps up around the point of scission, as translational motion does not set in until after scission occurs. As we show later, this significant internal collective excitation energy corresponds to a process where large-amplitude oscillations in shape occur. For the DIF case examined in Fig. 10 of Ref.~\cite{Goddard2015}, the collective energy was much smaller up until around the point of scission. This is an indication that the pre-scission configuration is physically different in the instantaneous BIF and the DIF processes. Whereas the latter corresponds to a relatively gentle pre-formation and scission mechanism, BIF is a more violent process. 

Differences also arise at the level of the kinetic energy of Fig.~\ref{symm-iso-fiss}(g). Whereas in the DIF case the total kinetic energy increased progressively by about $200$ MeV up until the point of fission, in the BIF case one observes two different features. Within the first $200$ fm/$c$, the kinetic energy oscillates by about $800$ MeV as the EDF accommodates the quadrupole boost. The kinetic energy then settles and oscillates around its initial value of $\approx 4400$ MeV with an amplitude of $\approx 100$ MeV. We take this as a sign of a quick rearrangement of the system within the reabsorption phase, followed by relatively milder large-amplitude oscillatory phase.

\begin{figure*}[t!]
\begin{center} 
\includegraphics[width=0.6\linewidth]{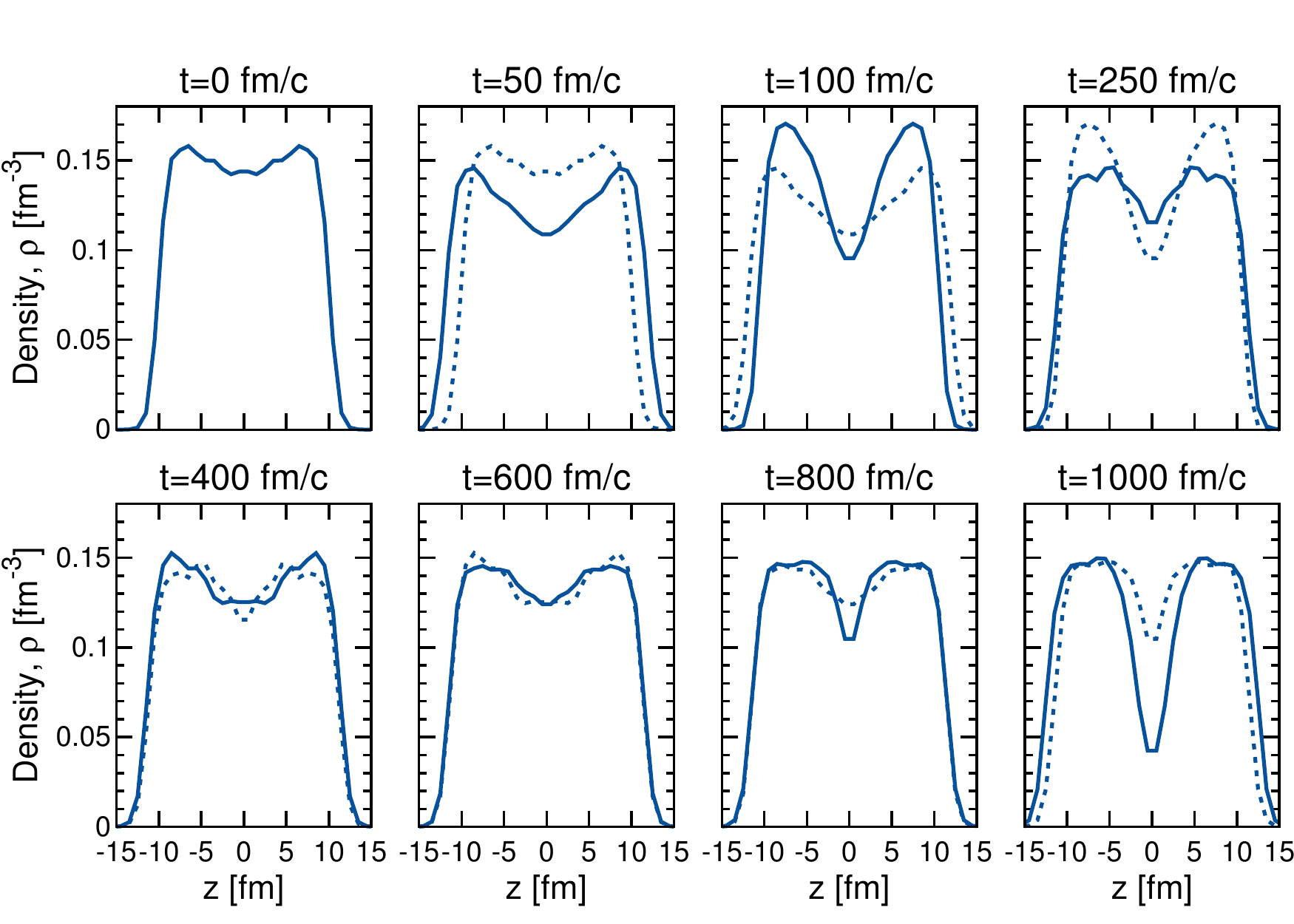} 
\caption{(Color online) One-dimensional slices of the particle density along the principal axis for various times, following the application of an instantaneous $200$~MeV quadrupole excitation upon the isomeric state. Densities at a specified time (solid lines) are compared to the density in the previous time (dotted line), to highlight the oscillatory nature of the pre-fission configuration. The final panel is very close to the scission point.}
\label{is_quad_boost_dens-slice}
\end{center}
\end{figure*} 

 Terms $E_0$-$E_3$ in Fig.~\ref{symm-iso-fiss}(a)-\ref{symm-iso-fiss}(d) also experience very large oscillations within the first $50$-$100$ fm/$c$. The rearrangement of the density-dependent terms $E_0$ [panel (a)] and $E_3$ [panel (d)] is much greater here than in the DIF case. When the system adjusts after about $\approx 200$ fm/$c$, the remaining oscillations are of the order of $\approx 200-300$ MeV.  Further, around the point of scission ($\approx 900-1000$ fm/$c$), there is a small increase in the average value around which the oscillations are based. By comparing the average of the oscillating values before and after fission, the $E_0$ and $E_3$ terms increase by $250$-$300$ MeV. The oscillatory nature of these terms make these values approximate, but they may be compared to the typical changes in magnitude observed in the DIF case at the point of scission, of $\approx 400$ MeV for the $E_0$ and $E_3$ terms. These differences may be attributed to the different final fission products for the BIF case as it is the local particle densities within the fragments that determine the postscission values of the $E_0$ and $E_3$ terms.

A similar picture arises when comparing the BIF $E_1$ [Fig.~\ref{symm-iso-fiss}(b)], $E_2$ [Fig.~\ref{symm-iso-fiss}(c)] and spin-orbit [Fig.~\ref{symm-iso-fiss}(e)] terms to their DIF counterparts, with an order-of-magnitude-larger energy variation in the immediate aftermath of the quadrupole boost and much milder oscillations after the system has absorbed the energy. The $E_2$ term for the fissioning case levels off at approximately $375$ MeV, in comparison to the nonfissioning case at $325$ MeV. For the fissioning case, we interpret the additional energy in terms of the nucleus necking as it fissions, which creates a larger surface. Around scission, the $E_1$ term increases by $\approx 15-25$ MeV. All in all, this indicates that the system goes through an initial violent phase of reconfiguration, followed by relatively milder large-amplitude oscillations that may or may not lead to fission, depending on the initial energy deposition and the oscillatory dynamics that follow. 

\begin{figure*}[t!]
\begin{center} 
\includegraphics[width=0.6\linewidth]{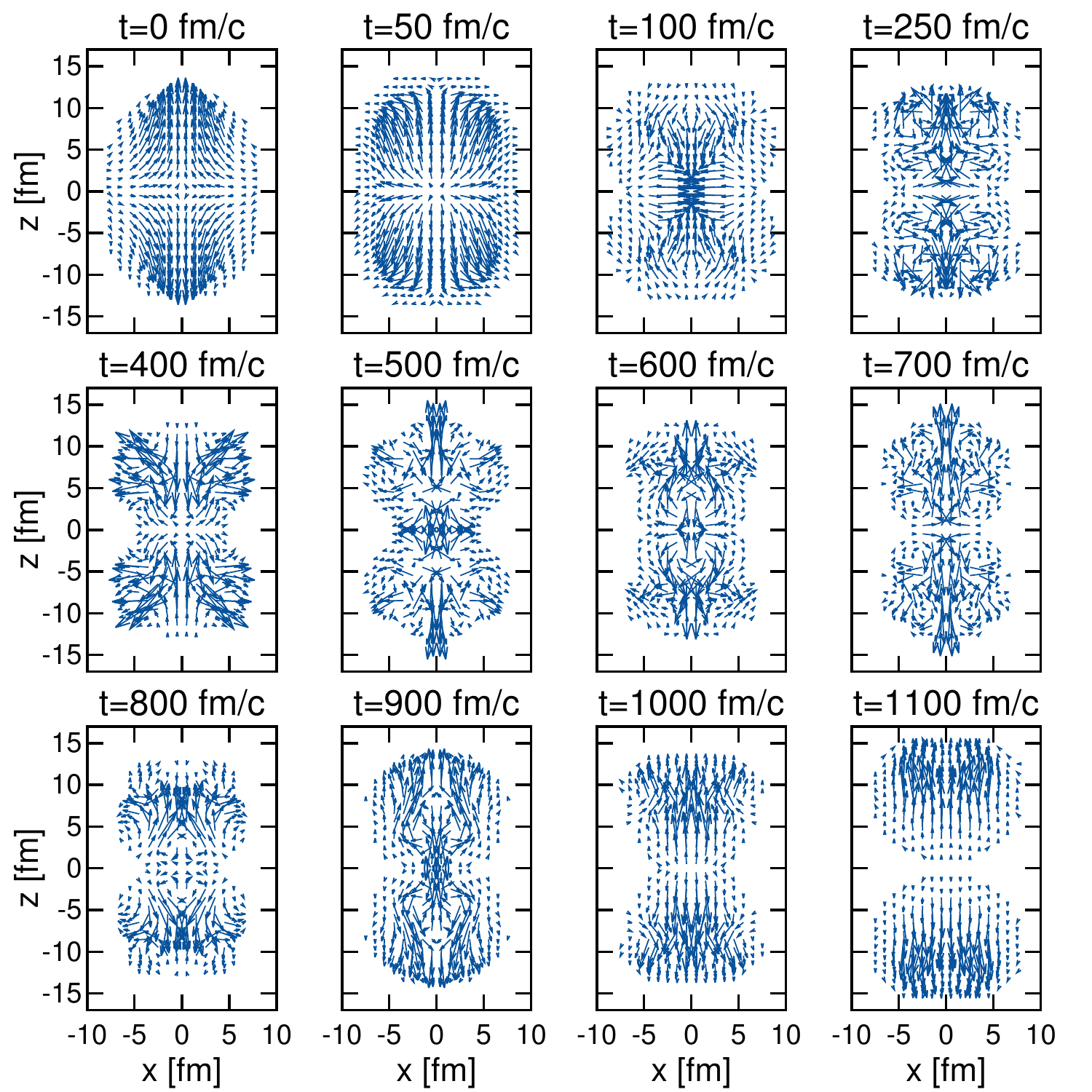} 
\caption{(Color online) Current vectors corresponding to the particle density slices presented in Fig.\@ \ref{is_quad_boost_dens}. The normalization factors for the vector arrows from $t=250$ to $t=900$ are the same (=$\mathcal N$), and for other times the normalization factor has been rescaled so that the panels are visually instructive. The normalization factors are $\mathcal N/10, 3\mathcal N/10$, $3\mathcal N/10$, $4\mathcal N/10$, and $3\mathcal N/10$ for $t=0,50,100,1000$ and $t=1100$ fm/$c$, respectively.}
\label{is_quad_boost_curr}
\end{center}
\end{figure*} 

As expected, the Coulomb contribution to the EDF ][Fig.~\ref{symm-iso-fiss}(f)] is very different for a final fissioning configuration (dot-dashed line) than for a nonfissioning one (dashed line). A compact nucleus does not show a decreasing Coulomb contribution with time, as two fission fragments separate from each other. In contrast to the other terms, the fissioning configuration of the Coulomb term provides a very similar picture to the DIF case. The Coulomb term is responsible for the scission process, which is qualitatively similar in both approaches. The Coulomb energy decrease is compensated by a large increase in collective kinetic energy as the fragments separate and accelerate. The overall energy is conserved within TDHF after the excitation has been applied, as shown in panel (j). The difference in total energy between the two boosted fragments reflects the $50$-MeV difference in the initial boost.

Figure \ref{is_quad_boost_dens-slice} displays 1D slices of the particle density along the principal axis of the nucleus for different times following the application of the $200$-MeV instantaneous excitation. These 1D plots are instructive when examined in conjunction with the 2D density slices presented in Fig.\@ \ref{is_quad_boost_dens}. Comparing the first two panels at $t=0$ fm/$c$ and $t=50$ fm/$c$, we find that the particle density follows an initial rapid elongation as the system is boosted by the quadrupole operator. This corresponds to the large initial change in absolute magnitude of the terms in the EDF (Fig.~\ref{symm-iso-fiss}). 
Following the initial expansion, the density is drawn back sharply at around $t=100$ fm/$c$ as the system recovers from the quadrupole stretch. The 1D slice of the particle density at this time displays a prominent dip around $z=0$, as a neck begins to develop. Subsequent oscillations of the particle density remain relatively compact in nuclear shape, but strain the neck further and further until fission becomes possible. 

Finally, the current vectors for the fissioning isomeric configuration are displayed in Fig.\@ \ref{is_quad_boost_curr}. They provide a useful visual aid when examined in conjunction with the particle densities. Following the initial stretching phase ($t=0$ to $t=50$ fm/$c$),  current vectors at $t=100$ fm/$c$ demonstrate necking occurring as the particle flow draws in at the neck region.  This coincides with the time at which the characteristic behavior of the hexadecapole deformation parameter corresponding to necking occurs, as seen in Fig.\@ \ref{is_quad_boost}(c). Further on, the current vectors reverse direction and the drawing in phase begins (between $t=50$ and $t=250$ fm/$c$). Beyond $250$ fm/$c$, the density gradually transitions into a fissioned configuration by means of a series of shape oscillations. 

The oscillations in the deformation  described in Fig.~\ref{is_quad_boost} are produced by a  ``sloshing" motion. After matter draws into the central region, an area with well-defined current vectors moving in phase but in opposite directions develops, leaving a static void behind. This is reminiscent of a shockwave and it is perhaps most clearly seen in the panel of Fig.~\ref{is_quad_boost_curr} corresponding to $t=600$ fm/$c$. The oscillatory nature of the deformation parameters (Fig.\@ \ref{is_quad_boost}) and decomposed terms of the energy functional (Fig.\@ \ref{is_quad_boost_dens-slice}) suggest that the initial shockwave occurs following the first decompression phase. Another shockwave occurs when the particle flow hits the central region, and reverses direction once more. This periodic sloshing effect is seen in the evolution of the deformation parameter and the energy functional, which oscillate with a characteristic period of $\approx 100$ fm/$c$. The behavior continues until $\approx 1000$ fm/$c$, where the sloshing has pulled the two prefragments to a point where the system can evolve into a fissioned configuration. Compared to the current densities presented for the DIF case in Ref.~\cite{Goddard2015}, the evolution of the current densities in instantaneous BIF is far more dramatic.

\subsection{Instantaneous BIF beyond the second barrier peak}

\begin{figure}[t!]
\begin{center} 
\includegraphics[width=0.6\linewidth]{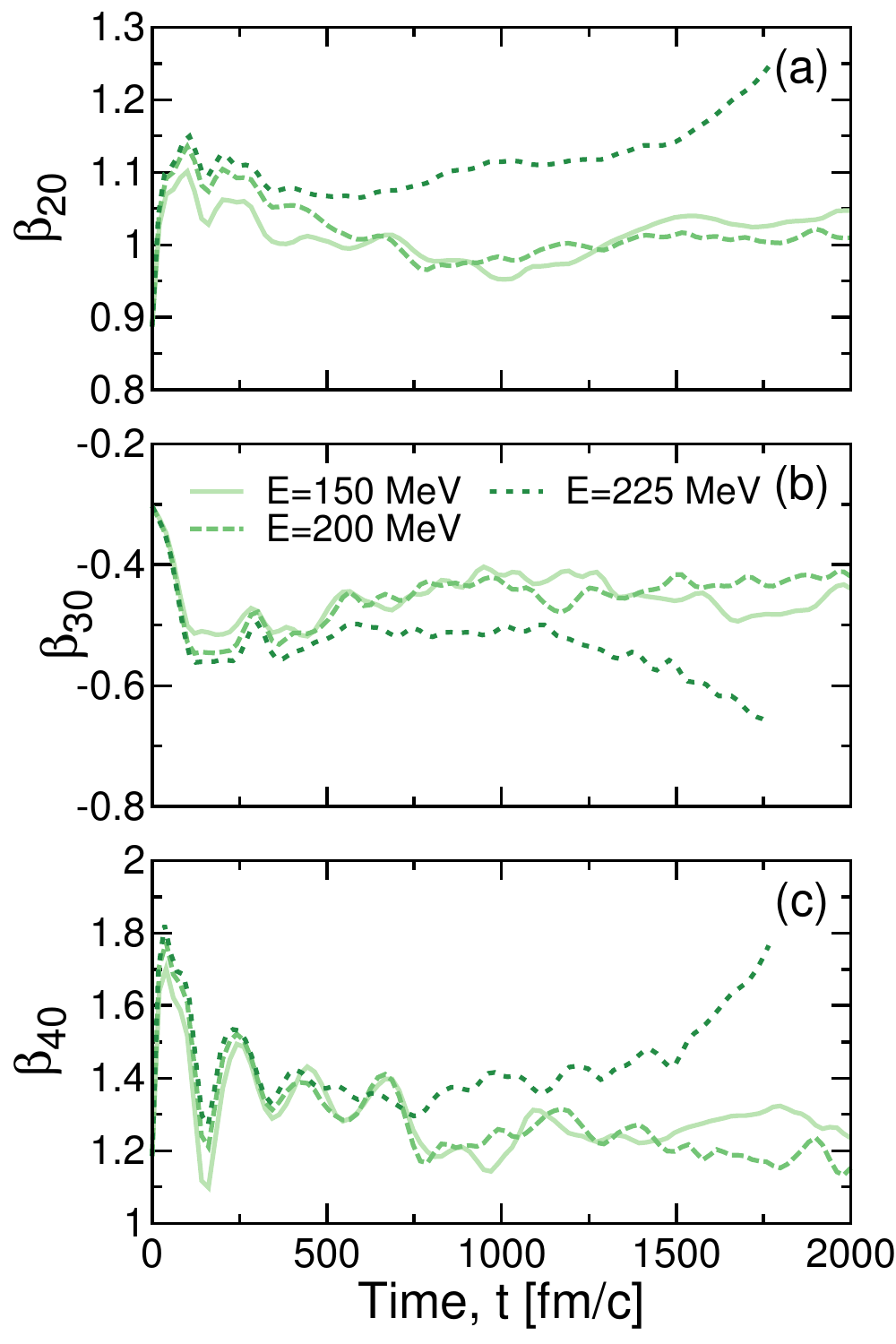}
\caption{(Color online) Time evolution of (a) quadrupole, (b) octupole, and (c) hexadecapole deformation parameters following a large-amplitude quadrupole excitation upon  the initial state with $\beta_{20}=0.890$. The threshold for fission is between $200$ and $225$ MeV. Scission occurs around $1700$ fm/$c$, and the measurements of the multipole deformation parameters are cut off at this point.}
\label{0_89_quad}
\end{center}
\end{figure} 

A similar investigation of BIF using instantaneous excitation fields may be considered, starting from the static state with quadrupole deformation $\beta_{20}=0.890$. The state lies just beyond the peak of the second static fission barrier, but fails to fission within an unboosted TDHF evolution of $9000$ fm/$c$. For this static state, mass asymmetry is present and octupole degrees of freedom are explored. Figure \ref{0_89_quad} shows the evolution of the multipole parameters following quadrupole excitations of various energies. Whereas the boosts with energies $150$ and $200$ MeV are insufficient to fission the system, a quadrupole excitation of $225$ MeV brings the system to fission, as evidenced by an increasing $\beta_{20}$ parameter. 

The threshold energy required to induce fission is thus in the range $200\le E_{\text{thresh}}\le 225$ MeV. This is $25$ MeV higher than the  boost required to fission the isomeric state (see Fig.\@ \ref{is_quad_boost}). This is a surprising result when considering the static PES. The initial configuration is more deformed ($\beta_{20}=0.890$) than the isomer ($\beta_{20}=0.682$), and hence one would expect that less energy should be required to induce fission. However, as previously mentioned, by applying an instantaneous boost, the state is removed from the static configuration and the corresponding PES. At $t=0$, despite the particle density being identical to the static configuration, the boosted state contains a large excitation in the form of collective kinetic energy. This highly excited state does not correspond to the static counterpart, and does not resemble anything encountered on the static PES. Moreover, for this state the boosted energy can be transferred into octupole deformation energy, which was not a possibility for the $\beta_{20}=0.682$ initial state.

The dynamics of the quadrupole degree of freedom are relatively similar to the isomeric-state excitation. For the nonfissioning configurations, a rapid increase in quadrupole within the first $100$ fm/$c$ is followed by a decrease in magnitude, towards the static value, but with substantial oscillations. Figure~\ref{0_89_quad}(b) shows the evolution of the octupole degree of freedom in time. In contrast to the isomeric state, here octupole deformations are actively explored. For all the cases, one finds that the octupole increases in absolute magnitude  from its original value $\beta_{30}=-0.3$ to $\beta_{30} \approx -0.55$. Whereas in the nonfissioning cases the octupole subsequently saturates and oscillates over time, the fissioning configuration leads to an increasing octupole deformation parameter, which reaches $\approx -0.7$ by the point of scission. Asymmetric fission fragments are produced, with scission occurring around $1700$ fm/$c$. 

\begin{figure*}[t!]
\begin{center} 
\includegraphics[width=0.6\linewidth]{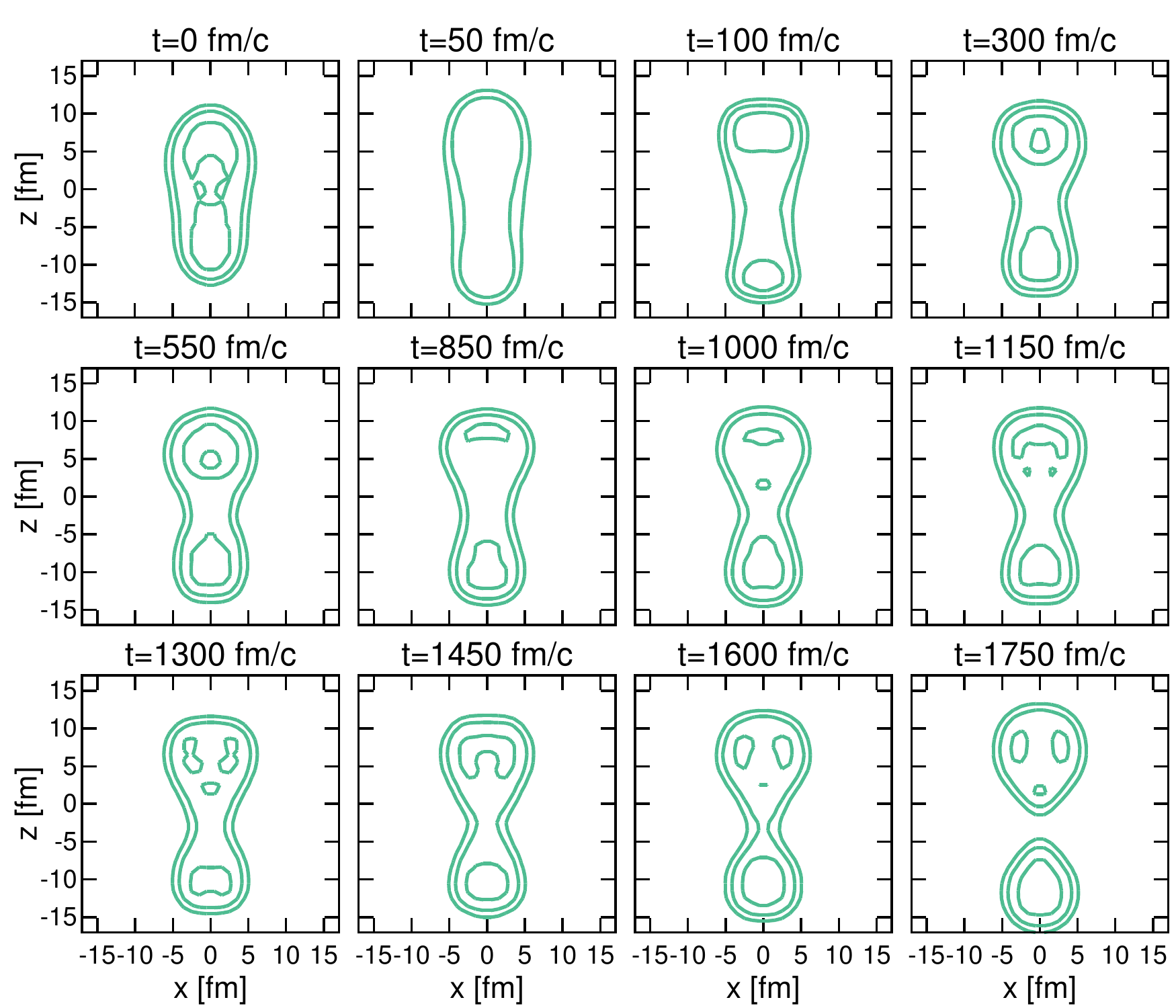}
\caption{(Color online) Slices of the total particle density for various times, following an instantaneous $225$ MeV quadrupole excitation upon the state with initial deformation $\beta_{20}=0.890$. The isolines are separated by $0.05$ particles/fm$^3$.}
\label{089_dens_slices}
\end{center}
\end{figure*} 

Figure \ref{089_dens_slices} shows the time evolution of the particle density following an instantaneous quadrupole excitation delivering $225$ MeV of energy. The initial state is already asymmetric in shape. The quadrupole boost immediately pulls apart two asymmetric lobes, connected by a low-density neck. This corresponds to the dip in hexadecapole moment [Fig.~\ref{0_89_quad}(c)] at about $100$-$200$ fm/$c$. Oscillations in the shape set in afterwards, with the system exploring an increasingly asymmetric configuration.  The increasing quadrupole parameter is associated with a growing elongation, with the width of the neck gradually shrinking. Ultimately, the system scissions into two asymmetric fragments.  

Table \ref{frag-ke-exp-tab} displays the mass numbers and energies of the fission fragments produced when applying the threshold instantaneous boosts required to observe BIF for the isomer and the state with static deformation $\beta_{20}=0.890$. BIF upon a mass symmetric isomer leads to mass-symmetric fission products, two excited $^{120}$Ag$_{47}$  isotopes. In contrast, application of the instantaneous excitation field to the state with initial deformation $\beta_{20}=0.890$ results in asymmetric fission products. In particular, the asymmetric mass fragments, which rounded to the nearest integer correspond to $^{151}$Pr$_{59}$ and $^{89}$Br$_{35}$, are about 30 mass units above and below the symmetric fragments. As expected from an isoscalar boost, all fission fragments have relatively similar $N/Z$ values.

\begin{table*}[t!]
\caption{Threshold energies and masses obtained from applying instantaneous excitation fields to the fission isomer. The masses for the minimum energy case observed to induce fission are presented. The extrapolated collective energy, corresponding mainly to translational kinetic energy, is computed using the extrapolation procedure detailed in Ref.~\cite{Goddard2015}.}
\centering 
\begin{tabular}{|c|c|c|c|c|c|c|} 
\hline 
 Static & $E_{\text{boost}} $& Heavy  & Light  &Heavy & Light&Extrapolated \\
State & [MeV] &Fragment&Fragment&Fragment&Fragment&Coll. KE \\
          &            & $(A,Z)$     &  $(A,Z)$         &         (Integer)          &  (Integer)& $[$MeV$]$ \\
\hline\hline 
    Isomer      &       200  & 120.00(5)      &     120.00(5)     &     $^{120}$Ag$_{47}$           &  $^{120}$Ag$_{47}$ & 218(8) \\ 
    & & 47.00(5) &  47.00(5)  & & & \\ \hline
     $\beta_{20}=0.890  $    &225&150.50(5) &89.49(5)   &$^{151}$Pr$_{59}$ &$^{89}$Br$_{35}$ &189(6) \\ 
        & & 58.78(5) &  35.23(5)  & & &\\ \hline
\end{tabular} 
\label{frag-ke-exp-tab}
\end{table*}

Column 7 of Table \ref{frag-ke-exp-tab}  also shows a figure for the collective kinetic energy of the system, which by and large dominates the energy balance of the outgoing fragments. 
The final collective kinetic energies of the fissioned systems have been deduced using the extrapolation procedure described in Ref.~\cite{Goddard2015}. The collective kinetic energy of the system is extrapolated using a Coulomb-like trajectory, and the final translational kinetic energy can thus be deduced. This assumes that the collective energy is dominated by translational kinetic energy, an assumption that has been checked by performing alternative calculations of the collective energies~\cite{Goddard2015}.

In both cases, the bulk of the excitation energy is absorbed into the nuclear terms of the energy functional within the first few hundred fm/$c$ (Fig.\@ \ref{symm-iso-fiss} is representative), and the remaining excitation is of the form of internal collective kinetic energy as a current is induced. Scission does not occur until well after the boost is applied, $t\approx1000-2000$ fm/$c$. To some extent, this demonstrates that it is not the boost itself that directly induces fission. Instead, the boost provides an onset of collective energy that is absorbed into shape excitations. If the shape oscillations are strong enough, they can lead to fission within a few periods. Because fission is not directly induced by the boost, the corresponding collective kinetic energies of the fragments are not in correspondence to the initial quadrupole boost energies. The mass distributions obtained for this state ($\beta_{20}=0.890$) by BIF will be compared to experimental data and DIF results in Sec.\@ \ref{sec:massdistbifdif}.

So far, the only cases that we have presented are very close to the energy threshold required for BIF. However, as we have just explained, in BIF there is not a one-to-one correspondence between the boost energy and the final fission fragments. Consequently, one can explore a variety of fission products by starting from the same initial state and boosting it with increasing energies beyond its BIF threshold. Table \ref{089_boosts} presents the fission products following more and more intense instantaneous quadrupole boosts. On the one hand, we find that a variety of (asymmetric) fission fragments are produced, within $2$ to $3$ mass units of the threshold fragments and very similar isospin content. On the other, we find that the collective kinetic energies of all these fragments are very close to each other, in a region of $\approx 180-190$ MeV. As a matter of fact,  the extrapolated collective kinetic energies agree within uncertainties for a charge difference of $\pm 1$ in the fission products. Again, this points towards the fact that large-amplitude shape oscillations are responsible for the fission process, rather than the boost itself. To some extent, the memory of the initial boost is not relevant for the final fission product kinetic energies. 

The results corresponding to the two extreme boosts, $E=225$ and $400$ MeV, are interesting in that the final fragments are (to the nearest integer) the same. This provides a verification of the assumption that most of the energy in the fission process goes into the collective kinetic fragment energies. Owing to the Coulomb interaction imparting the translational kinetic energy, the resulting values should agree if the contribution from internal collective excitations are small. The resulting extrapolated energies do, in fact, agree within uncertainties, demonstrating that the energy released in BIF is still dominantly translational kinetic energy, even for boost energies beyond the threshold for fission. 

\begin{table*}[t!]
\caption{Fission products obtained by BIF, applying an instantaneous quadrupole excitation of various energies to the state with initial deformation $\beta_{20}=0.890$. The collective kinetic energy, corresponding mainly to translational kinetic energy, is computed using the extrapolation procedure detailed in Ref.~\cite{Goddard2015}.}
\centering 
\begin{tabular}{|c|c|c|c|c|c|} 
\hline 
 Boost      &  Heavy        & Light                &Heavy                   & Light&Extrapolated \\
Energy    &Fragment    &Fragment         &Fragment              &Fragment&Coll. KE \\
 $[$MeV$]$       & $(A,Z)$       &  $(A,Z)$           &         (Integer)       &  (Integer)&$[$MeV$]$ \\
\hline\hline 
225 &150.50(5) , 58.78(5)&89.49(5) , 35.23(5) &$^{151}$Pr$_{59}$&$^{89}$Br$_{35}$&189(6) \\ \hline
250 &147.61(5) , 57.71(5)&92.47(5) , 36.28(5) &$^{148}$Ce$_{58}$&$^{92}$Kr$_{36}$&189(2) \\ \hline
300 &147.06(5) , 57.50(5)&92.92(5) , 36.50(5) &$^{147}$Ce$_{58}$&$^{93}$Rb$_{37}$&188(4) \\ \hline
350 &148.37(5) , 58.10(5)&91.62(5) , 35.90(5)&$^{148}$Ce$_{58}$&$^{92}$Kr$_{36}$&180(3) \\ \hline
400 &150.61(5) , 58.51(5)&89.37(5) , 35.48(5)&$^{151}$Pr$_{59}$&$^{89}$Br$_{35}$&176(11) \\ \hline
\end{tabular}
\label{089_boosts}
\end{table*}

\begin{figure}[h!]
\begin{center} 
\includegraphics[width=\linewidth]{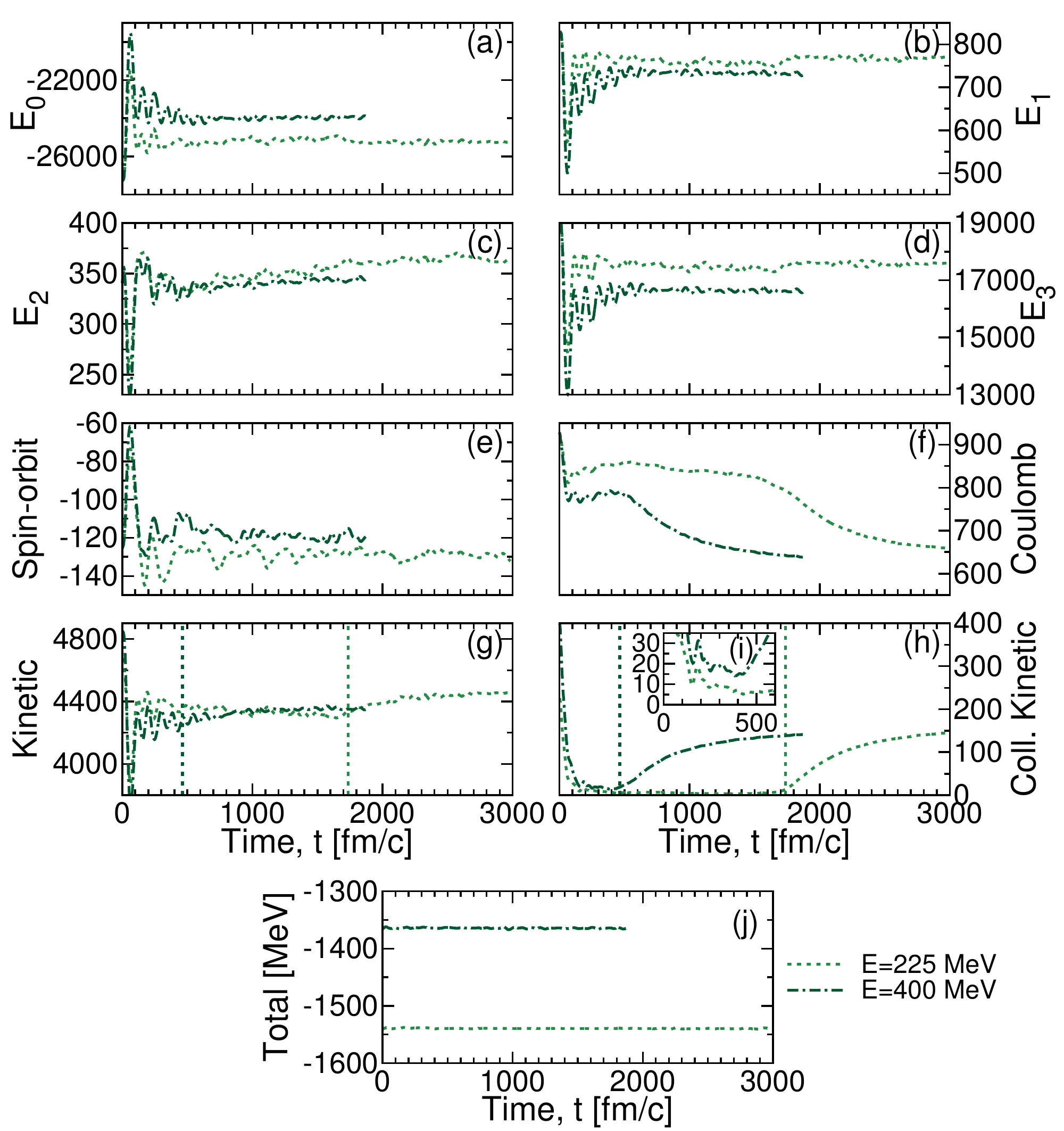} 
\caption{(Color online) (a)-(j): Evolution of the integrated contributions to the EDF following instantaneous quadrupole excitations of the state with initial deformation $\beta_{20}=0.890$. The dashed line corresponds to a nonfissioning state, whereas the dash-dotted line represents a fissioning configuration. Vertical lines in panels (g) and (h) correspond to the point of scission. Units in all panels are MeV. See text for more details.}
\label{089-kicks}
\end{center}
\end{figure} 

The time evolution of the different terms of the EDF are a useful tool to analyze the dynamics of the BIF process. These are shown for the  $225$ (dotted lines) and $400$ MeV (dot-dashed line) excitations applied to the state with $\beta_{20}=0.890$ in Fig.~\ref{089-kicks}. The evolution of the energy components displays two very different time scales for fission. The point of scission is marked with vertical lines on panels (g) and (h). Whereas scission occurs within $450$ fm/$c$ for the $400$-MeV excitation, it takes approximately $1650$ fm/$c$ for the state excited with $225$ MeV to fission. In both cases, the bulk of the initial excitation energy is absorbed within the first $100$ fm/$c$, and quickly transferred into the nuclear parts of the EDF. The initial oscillations in the nuclear terms of the EDF are similar for both cases within the first $200$ fm/$c$. The amplitude of the oscillations is noticeably larger for the $400$ case, though, indicating more important shape reconfiguration processes.

\begin{figure*}[t!]
\begin{center} 
\includegraphics[width=0.6\linewidth]{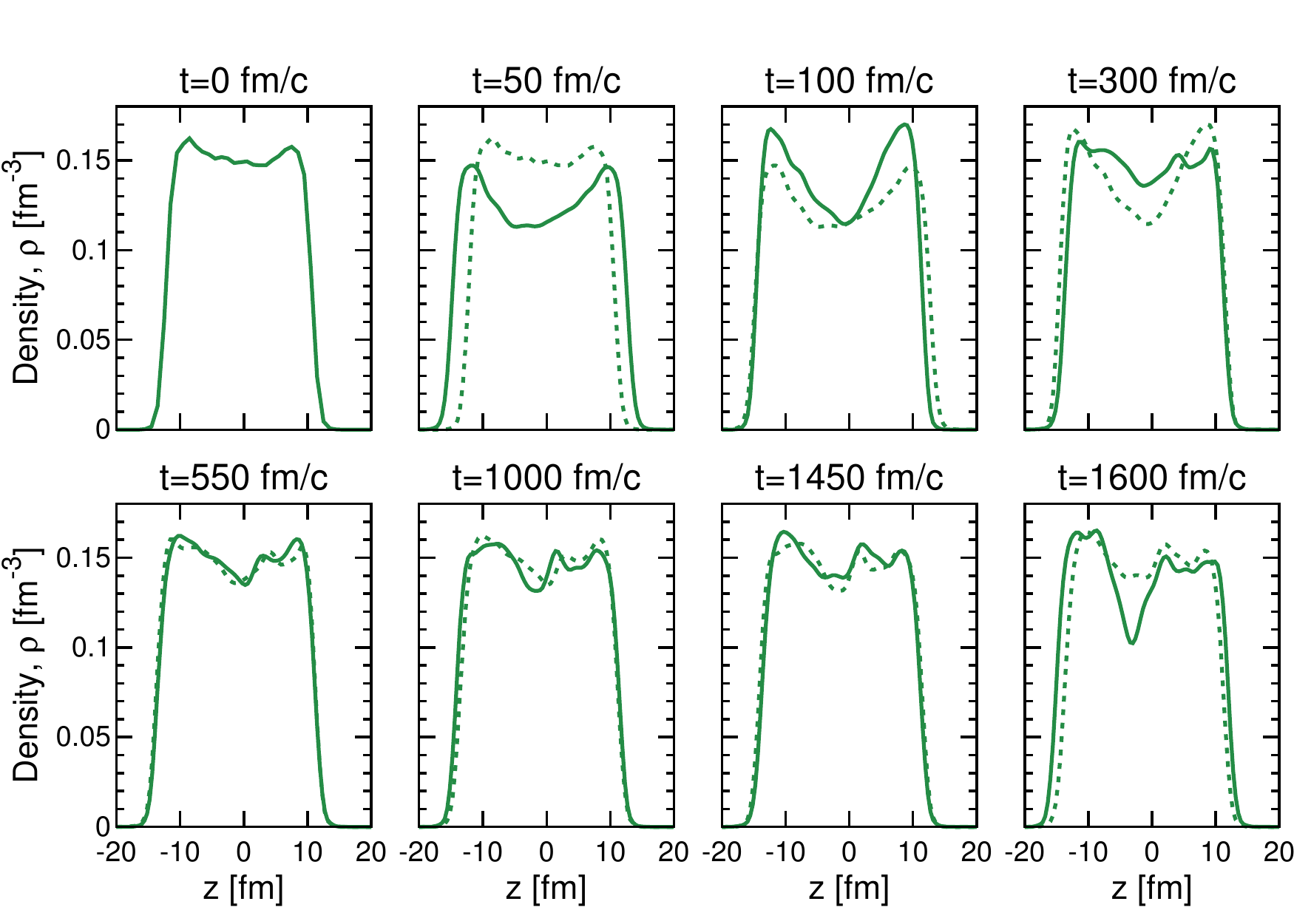} 
\caption{(Color online) One-dimensional slices of the particle density along the principal axis for various times, following the application of an instantaneous $225$~MeV quadrupole excitation upon the state with initial deformation  $\beta_{20}=0.890$. Densities at a specified time (solid lines) are compared to the density in the previous time (dotted line).}
\label{089-kick-225-1d}
\end{center}
\end{figure*}

\begin{figure*}[t!]
\begin{center} 
\includegraphics[width=0.6\linewidth]{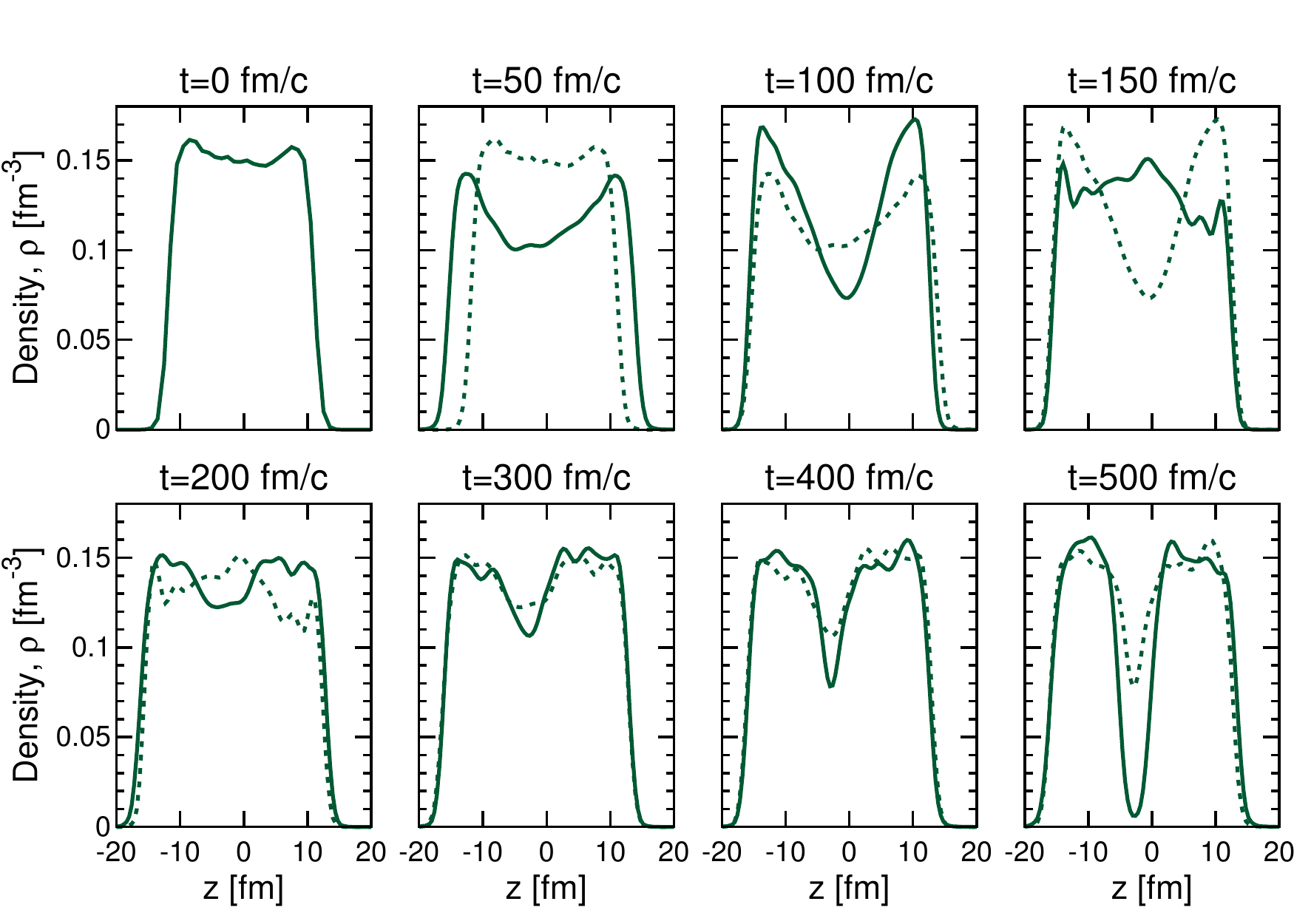} 
\caption{(Color online) The same as Fig.~\ref{089-kick-225-1d} for an excitation of $E=400$ MeV.}
\label{089-kick-400-1d}
\end{center}
\end{figure*}

To explain the differences in the fission timescales when applying the two boosts, it is helpful to examine the snapshots of 1D slices of the particle densities. These are displayed in Figs. \ref{089-kick-225-1d} and \ref{089-kick-400-1d} for the $225$- and $400$-MeV excitations, respectively. For the $225$-MeV excitation, where a longer fission timescale is observed, the behavior is similar to that where a $200$-MeV excitation was applied to the isomer (Fig.~\ref{is_quad_boost_dens-slice}). Following the application of the boost, the nucleus is stretched, and then draws sharply back in within the first $100$ fm/$c$. By $300$ fm/$c$, virtually all of the excitation energy has been absorbed, and the remaining collective energy corresponds to the induced current. This corresponds to the $\approx 5$ MeV of collective kinetic energy that remains in the system as it rearranges [see Figs.~\ref{089-kicks}(h) and \ref{089-kicks}(i)]. From about $300$ fm/$c$ on, the density has recovered in the central region, and it is here that the shockwave behavior sets in, as seen previously for the isomer. During this phase, the densities slosh around as the particle flow travels outwards, then sharply reverses direction, and continues oscillating. The sloshing causes relatively irregular oscillations in the $E_0$-$E_3$ terms of the EDF [Figs.~\ref{089-kicks}(a)-\ref{089-kicks}(d)]. Beyond $1600$ fm/$c$, the particles in the neck have mostly transitioned into the two lobes, and the Coulomb repulsion drives the configuration to scission [panel (h)].

When applying the $400$-MeV excitation, a much faster fission timescale is observed. Figure \ref{089-kick-400-1d} displays the corresponding 1D density slices as the system evolves. As observed in the evolution of the energy functional (Fig.\@ \ref{089-kicks}), the amplitude of the oscillations in the decomposed terms for the first $500$ fm/$c$ are much larger than the $225$-MeV case. This corresponds to more significant oscillations and currents induced by the stronger excitation. The collective kinetic energy, after the initial absorption phase, only drops to $\approx20$ MeV [Fig.\@ \ref{089-kicks}(i)], in comparison to $\approx5$ MeV for the $225$-MeV boost. Moreover, we find faster and larger-amplitude oscillations in the EDF terms in panels (a)-(d) within the first $500-750$ fm/$c$. This allows the configuration to rearrange more rapidly. Indeed, the initial state requires far fewer oscillations of the particle flow before the nucleus rearranges such that the Coulomb repulsion drives the configuration into two fragments. Here, scission occurs between $400$ and $500$ fm/$c$. It is interesting to observe the differences in the $E_0$-$E_3$ and spin-orbit terms in Fig.~\ref{089-kicks} once the systems have fissioned. This suggests that the final fragments have different deformations (that is, the particle density is arranged differently), despite having the same $N$ and $Z$. This difference in shape configuration results from the trajectory followed to fission. Presumably, the fission products will be excited in different energy modes. While we do not carry out this analysis here, we note that we could explicitly analyze these excitation modes using the techniques developed in Ref.~\cite{Goddard2015}.

There is no particular reason to choose an upper excitation energy limit of $400$ MeV. We have actually experimented with more energetic boosts and, while we do not show further results here, we have found interesting results. In particular, an $E=800$ MeV boost produces a ternary fission event \cite{Goddard14,Tanimura2015}. An investigation of this fission mode would require a straightforward modification of {\sc sky3d} to incorporate three-fragment analysis, and may be of interest for future work.

\section{Time-Dependent Boosts}
\label{sec:ext}

As it has been demonstrated, the threshold excitation to observe BIF with an instantaneous boost requires an energy deposition of the order of $200$ MeV. As the energy is all deposited instantaneously, the correspondence between the static configuration and the state which is time-evolved is distorted. Adding energy to the system gradually may give the densities time to smoothly evolve into a fissioned configuration, in a manner comparative to DIF. In this section, the external excitation will thus be applied gradually via a time-dependent profile. Again, we consider both the isomer and the configuration with $\beta_{20}=0.890$ as initial states. 

The single-particle Hamiltonian $\hat{h}_q$ acting on the proton and neutron states can be modified to include the time-dependent isoscalar external field, $U_{\text{ext.},q}(\bm r , t)$ \cite{Mar13}:
\begin{equation}
\hat{h}'_q(t) = \hat{h}_q(t) + U_{\text{ext.},q}(\bm r , t) \, .
\label{eq:hprime}
\end{equation} 
Here, the external field $U_{\text{ext.},q}(\bm r , t)$ is given by
\begin{equation}
U_{\text{ext.},q}(\bm r , t) = \eta\, f(t)\, \phi_q(\bm r) \, .
\label{eq:uprime}
\end{equation} 
$\phi_q(\bm r)$ is the (quadrupole) spatial profile of the external field. The constant $\eta$ tunes the amount of energy added to the system. We note that time-dependence precludes us from finding a closed form for the energy as a function of $\eta$, unlike the instantaneous case (see the Appendix). We note, however, that the total energy is affected by the inclusion of this external field. One can thus read the total amount of deposited energy by monitoring the total energy of the system. The temporal profile of the excitation field is characterized by the functional $f(t)$, which we choose to be of Gaussian form:
\begin{equation}
\label{timeprof}
f(t) = \exp\frac{-(t-\tau_0)^2}{\Delta \tau^2} \, .
\end{equation}
The profile is centered around $\tau_0$, and has a width $\Delta \tau$. Values of $\tau_0$  in the region of $150$-$800$ fm/$c$ will be investigated, and $\Delta \tau$ will be taken as approximately $\frac{\tau_0}{3}$. 

\subsection{Time-dependent BIF on the isomeric state}
\label{isomer-tdep-sec}

\begin{figure}[h!]
\begin{center} 
\includegraphics[width=0.6\linewidth]{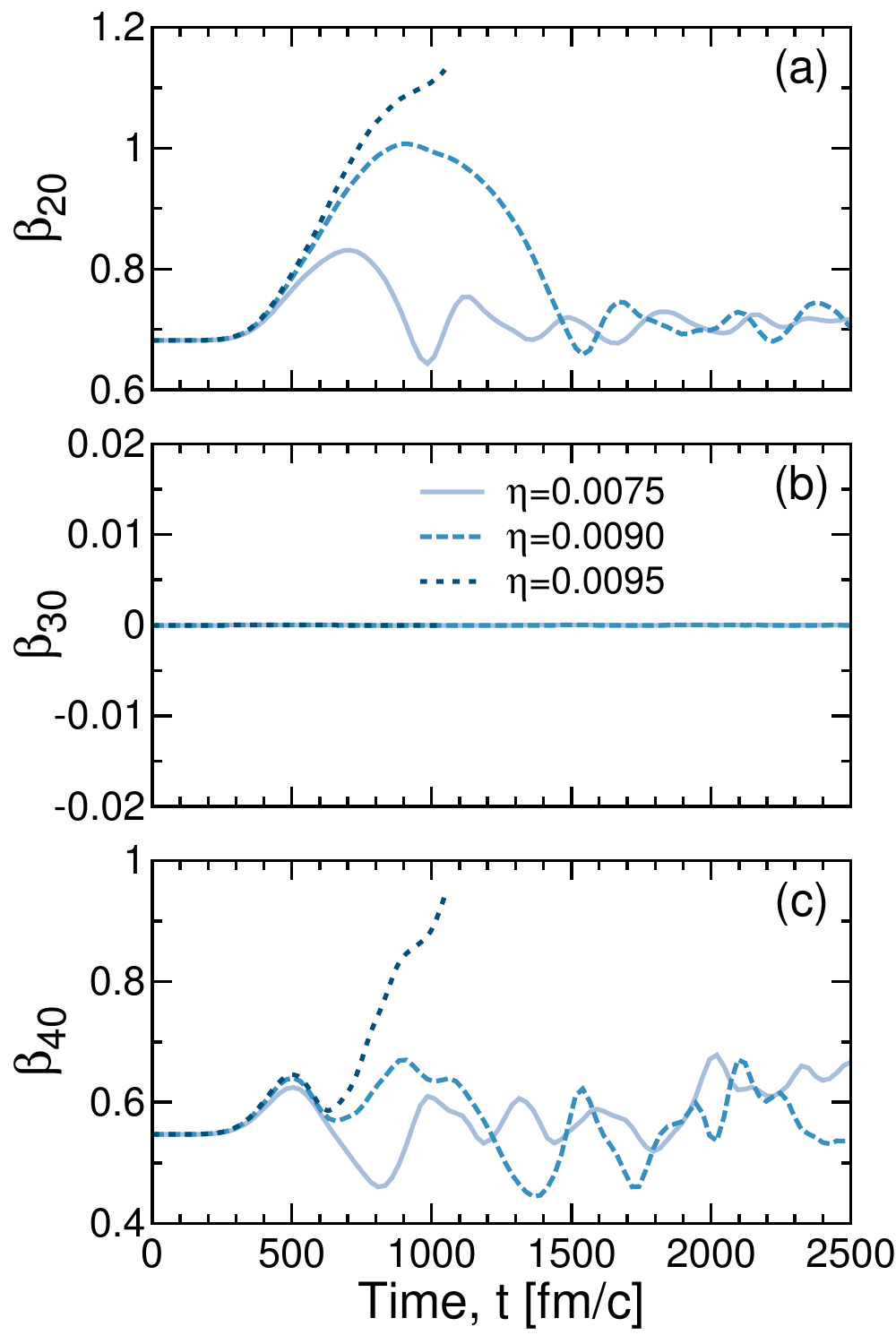} 
\caption{(Color online) Time evolution of (a) quadrupole, (b) octupole and (c) hexadecapole deformation parameters following time-dependent ($\tau_0=500$ fm/$c$ and $\Delta \tau=150$ fm/$c$)  quadrupole excitations upon the isomeric state. The field with scaling parameter $\eta=0.0095$ is seen to induce fission, and the evolution of the multipole moments are sharply cut off at the point of scission at $1050$ fm/$c$.}
\label{timedep-isomer-mpole}
\end{center}
\end{figure}

We start our discussion with an initial choice of $\tau_0=500$ fm/$c$ and $\Delta \tau=150$ fm/$c$ to describe the temporal profile of the external field. Figure \ref{timedep-isomer-mpole} displays the evolution of the multipole moments subject to an external field with increasing strengths, $\eta$. The two lowest values, $\eta=0.0075$ (solid line) and $0.0090$ (dashed line) do not produce a fissioning state, whereas $\eta=0.0095$ (dotted line) does. For the latter, the evolution of the multipole moments have been sharply cut off at the point of scission. In the cases where the nucleus fails to fission, the quadrupole deformation reverts back to the original value once the external excitation ends. Oscillations in the quadrupole [panel (a)] and hexadecapole [panel (c)] degrees of freedom are visible beyond this, and they are of the same order of magnitude as the instantaneous BIF case. The fissioning configuration is, in contrast, reached with a constantly increasing quadrupole degree of freedom and within one hexadecapole oscillation. 

\begin{figure*}[t!]
\begin{center} 
\includegraphics[width=0.6\linewidth]{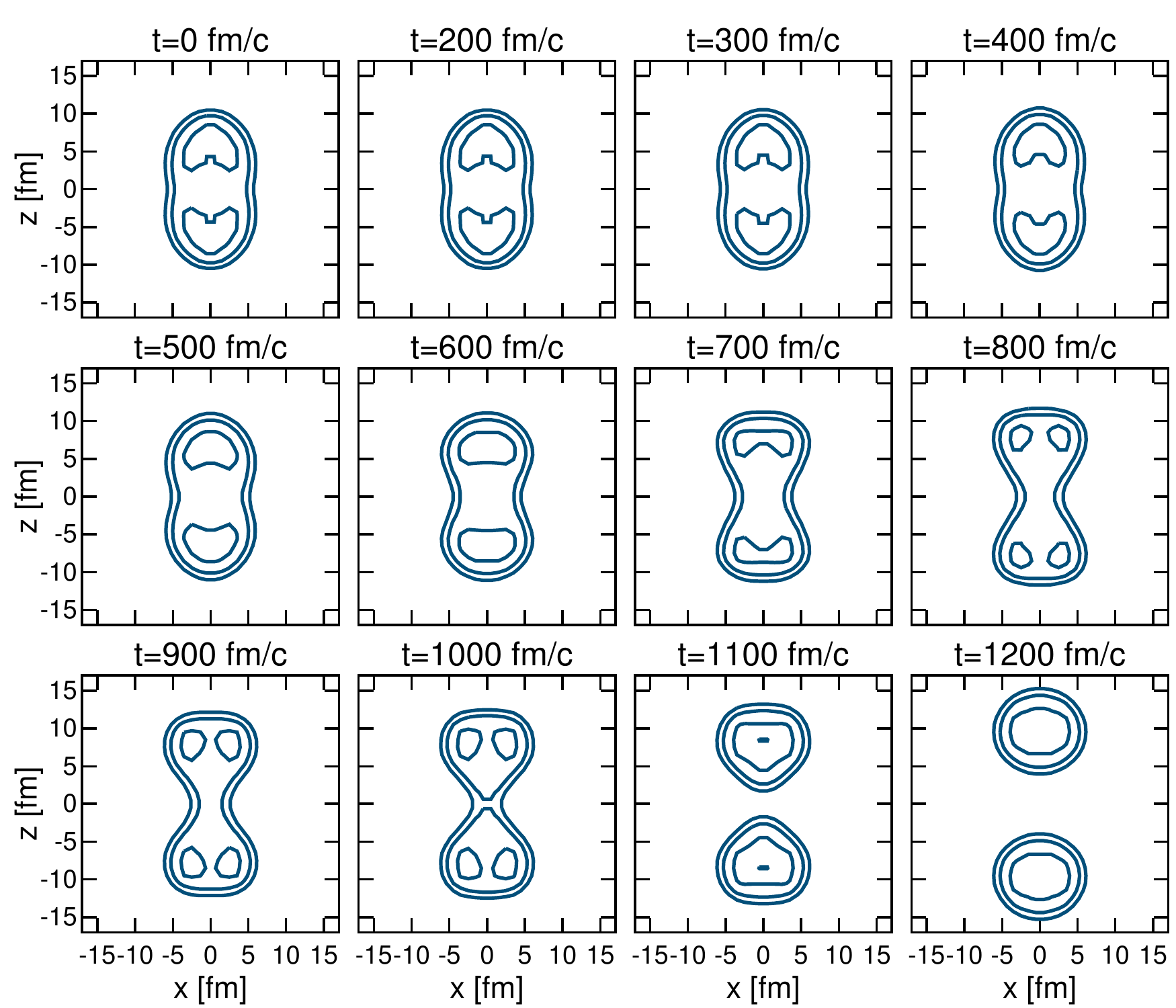} 
\caption{(Color online) Slices of the total particle density for various times, following a time-dependent boost with scaling constant $\eta=0.0095$ upon the isomeric state. The isolines are separated by $0.05$ particles/fm$^3$.}
\label{timedep-isomer-dens}
\end{center}
\end{figure*}

Figure \ref{timedep-isomer-dens} displays 2D slices of the particle density for the fissioning case ($\eta = 0.0095$). Symmetric fission into two $^{120}_{47}$Ag fragments is observed, as expected for a symmetric excitation to a symmetric system.  When applying the time-dependent excitation field, however, we only find visible deviations from the initial density after $\approx 400$ fm/$c$, whereas Fig.\@ \ref{is_quad_boost_dens} showed a dramatic immediate change in the nuclear configuration following the instantaneous excitation.

The hexadecapole degree of freedom also evolves differently here. There is no sharp drop in the value of $\beta_{40}$ [see Fig.~\ref{timedep-isomer-mpole}(c)], which indicates that the neck does not develop in the same way that it did in the instantaneous BIF. Instead, the nucleus elongates, thus increasing its $\beta_{20}$ value, and smoothly vibrates around the neck region. We note, however, that the timescale for scission is comparable to the instantaneous BIF case, as the system requires $\approx 1050$ fm/$c$ to fission. 

\begin{figure*}[t!]
\begin{center} 
\includegraphics[width=0.6\linewidth]{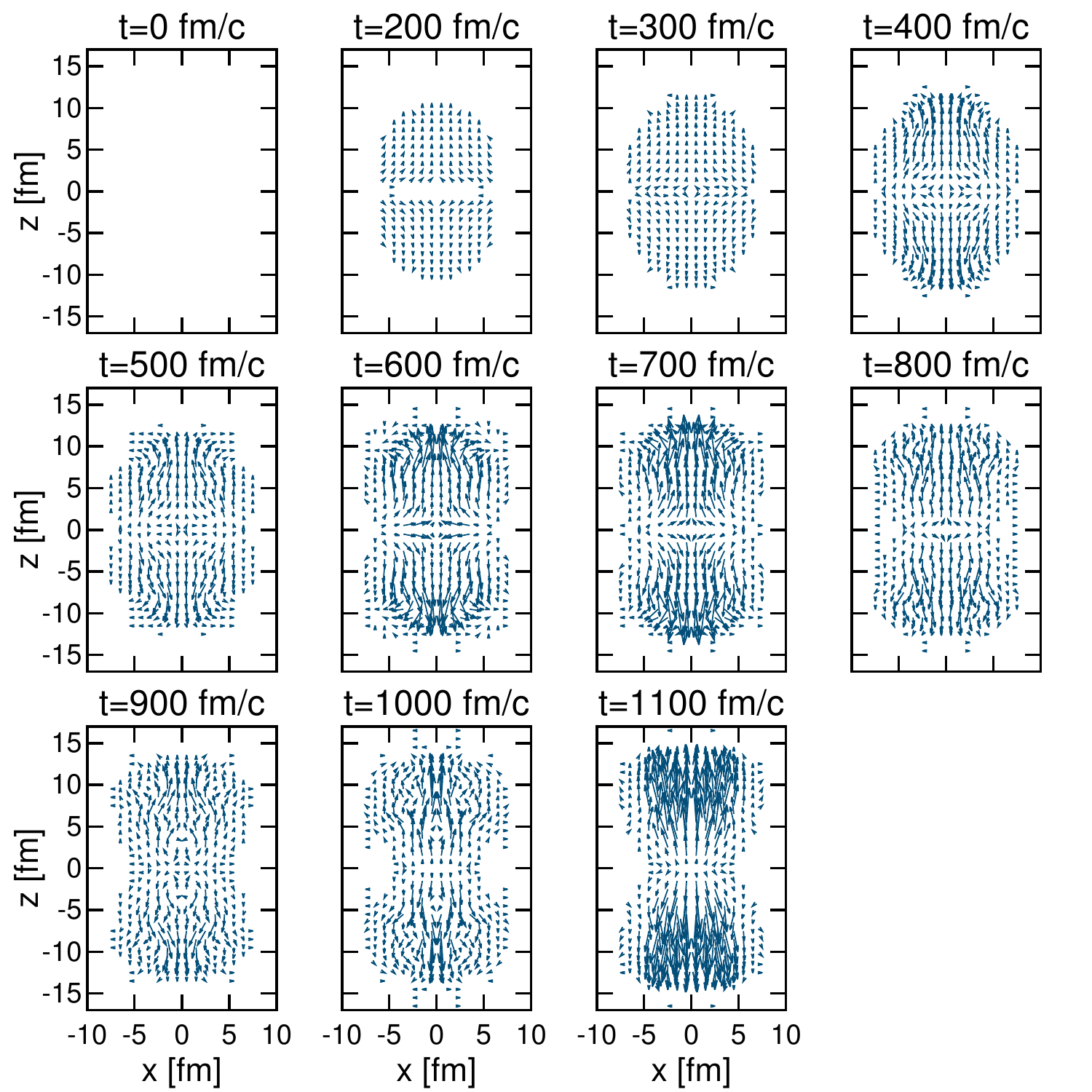} 
\caption{(Color online) Current vectors corresponding to the particle density slices presented in Fig.\@ \ref{timedep-isomer-dens}. The external field peaks at $\tau_0=500$ fm/$c$ with width $\Delta\tau$=150 fm/$c$, so the excitation field has reduced to a negligible magnitude by $t=800$ fm/$c$. The vectors have been normalized to be visually instructive. The normalization factor ($\mathcal N$) is the same in each panel, other than $t=1100$ fm/$c$, where it is $3\mathcal N/5$.}
\label{timedep-isomer-curr}
\end{center}
\end{figure*} 

The current vectors corresponding to the particle density slices presented in Fig.\@ \ref{timedep-isomer-dens} are shown in Fig.\@ \ref{timedep-isomer-curr}. The observed behavior may be compared to the DIF example in Ref.~\cite{Goddard2015} and the instantaneous BIF case presented earlier.  Figure \ref{timedep-isomer-curr} shows current vectors pointing outwards into the two fission fragments, gradually increasing in magnitude up to about $800$ fm/$c$, where the external field becomes negligible. In the DIF case, the currents in the two forming fragments pointed in opposite directions, with little contribution from the neck region. Here there are initially many more particles in the neck (the initial configuration is less deformed), resulting in a significant particle flow in this region, especially around $600$-$700$ fm/$c$ (Fig.\@ \ref{timedep-isomer-curr}). 

Beyond $800$ fm/$c$, a strong-enough current has been induced, and the nucleus has reconfigured itself so that it evolves to fission without further influence from the excitation field. It will be shown that the collective energy corresponding to this current is small compared to the instantaneous BIF case. The system continues evolving into a two-fragment configuration without the flow of particles begin drawn back inwards, similarly to the behavior seen for the DIF case. In contrast, the BIF case displayed significant oscillations and shockwave-type behavior. Unlike instantaneous BIF, there is no significant sloshing motion or oscillations in the density during the evolution to fission. This suggests that a physically different transition to the fissioned state is occurring for temporally extended BIF, similar to that seen for DIF. 

\begin{figure}[h!]
\begin{center} 
\includegraphics[width=\linewidth]{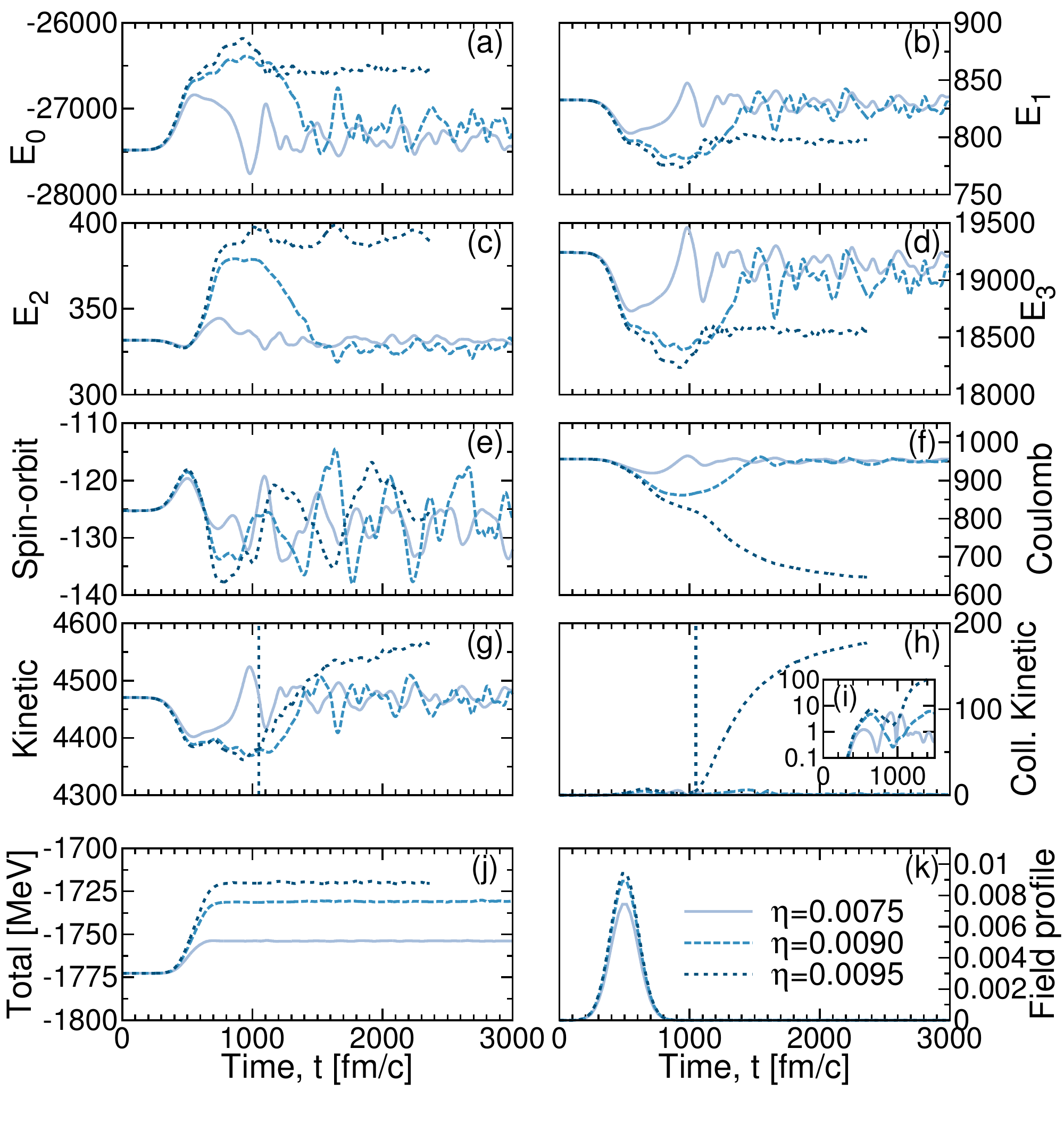} 
\caption{(Color online) (a)-(j): Time evolution of the integrated contributions to the EDF when applying a time-dependent external field to the isomeric state. The time profile of the external field is displayed in panel (k). The case with scaling constant $\eta=0.0095$, which corresponds to an excitation of $\approx 52$ MeV, is seen to fission. For this case, the calculation is terminated once the fragment separation exceeds $100$ fm. Vertical lines in panels (g) and (h) correspond to the point of scission. Units in all panels are MeV. See text for more details.}
\label{timedep-isomer-edf}
\end{center}
\end{figure}

It is instructive to examine the contributions to the EDF for all three time-dependent excitation fields presented in Fig.\@ \ref{timedep-isomer-mpole}. We provide these in Figs. \ref{timedep-isomer-edf}(a)-\ref{timedep-isomer-edf}(i). The temporal profile of the external field is shown, as a visual aid, in panel (k). The energy added to the system by the field for the different scaling parameters $\eta$ can be read off panel (j), corresponding to the total energy. As expected, time-dependent boosts with larger $\eta$ values deposit more energy into the system, with $\eta=0.0075$ injecting $E \approx20$ MeV; $\eta=0.0090$, $E \approx41$ MeV and $\eta=0.0095$, $E\approx52$ MeV.  It is remarkable to observe that fission has been induced by adding just $52$ MeV of energy. A threshold of $41\le E_{\text{thresh}}\le 52$ for this particular set of time-dependent parameter may be deduced. This compares to the threshold of $175\le E_{\text{thresh}}\le 200$ MeV for the instantaneous boosts. When applying a time-dependent excitation field, energy is deposited into the system in a more gradual manner, the system reconfigures in the quadrupole degree of freedom as the boost is applied and fission is reached more easily. We conclude that not only energy deposition, but also the time scale for the energy deposition, matters in terms of fission dynamics.

The time evolution of the contributions to the EDF shown in Fig.\@ \ref{timedep-isomer-edf} may be compared to the case of the instantaneous BIF presented in Fig.\@ \ref{symm-iso-fiss}. Some terms remain similar, particularly for the fissioning cases. For example, upon scission, the Coulomb and collective kinetic energies display behavior characteristic of two repulsively charged fragments accelerating away from one another [panels (f) and (h) of both figures]. In contrast, the evolution of the $E_0,E_1,E_2, $ and $E_3$ of panels (a)-(d) are drastically different. For the instantaneous BIF case  of Fig.\@ \ref{symm-iso-fiss}, all four of these terms were seen to display a prompt reduction in magnitude during the first $50-100$ fm/$c$ as the configuration underwent a rapid elongation. Following this, the magnitude of the terms recovered as the particle density evened out, and oscillated violently as shockwaves in the evolution of the densities set in. 

For the time-dependent BIF case (dotted line on Fig.\@ \ref{timedep-isomer-edf}), in contrast, the evolution of the energy functional displays a behavior which is more qualitatively similar to the DIF case of Ref.~\cite{Goddard2015}. The $E_0,E_1,$ and $E_3$ terms gradually decrease in magnitude and saturate, while oscillating, before scission occurs at $900$-$1100$ fm/$c$. Small oscillations about an approximately constant value go on in the postscission phase. These correspond to the collective excitations of the fission fragments. For the case of the instantaneous boost, the $E_2$ term in Fig.\@ \ref{symm-iso-fiss}(c) shows an initial reduction in magnitude of approximately $100$ MeV. In contrast, for the time-dependent excitation the evolution of the $E_2$ term shows only a dip of  $\approx5$ MeV in magnitude at around $500$ fm/$c$ as the shape configuration starts changing, before rapidly increasing by approximately $50$ MeV as the surface of the nucleus increases.

The absolute variation in the EDF terms for the time-dependent boosts are small compared to the instantaneous boosts, and they are generally closer to that observed for DIF. For example, the $E_0$ term in Fig.\@ \ref{timedep-isomer-edf} for the fissioning case changes by about $1500$ MeV. For the instantaneous BIF case, the change was about three times larger, $\approx5000$ MeV, whereas in DIF the maximum variation of the $E_0$ term during time evolution was in the region of $400$ MeV \cite{Goddard2015}.

The collective kinetic energy displayed in panel (h), and the inset panel (i), of Fig.\@ \ref{timedep-isomer-edf} is negligible up to $\approx 300$ fm/$c$. An initial peak appears just beyond the maximum of the external field time profile. It is interesting to observe that the collective kinetic energy reduces after the external field peaks, before scission occurs. For the fissioning case, at approximately $1000$ fm/$c$, the collective kinetic energy increases only as the system transitions into a fissioned configuration. This behavior is, again, similar to that seen for the DIF cases examined in Ref.~\cite{Goddard2015}. This suggests that the time-dependent external field has induced small internal currents, but it has gradually transitioned the nucleus into a shape configuration where fission becomes energetically favorable. In both BIF cases, whether instantaneous or time-dependent, the excitation energy from the boost is dissipated mainly into the nuclear terms in the functional. The induced current, corresponding to the collective energy up until around the point of scission, is small in comparison to the total excitation energy added to the system. In particular, no shockwave-type behavior is observed in the time-dependent BIF case as the energy is slowly released into the nucleus and a gradual evolution of the densities occurs.

The collective kinetic energy may be extrapolated using the procedure described in Ref.~\cite{Goddard2015}. We find that $221(1)$ MeV are released in the fission process, mainly owing to the translational kinetic energy of two symmetric $^{120}$Ag$_{47}$ fragments. This is in good agreement to the instantaneous BIF value, $218(8)$ MeV, of Table \ref{frag-ke-exp-tab}. This agreement is unsurprising considering that the fission products are identical, and it is the Coulomb interaction which imparts most of the final collective energy to the system.

\begin{figure}[h!]
\begin{center} 
\includegraphics[width=\linewidth]{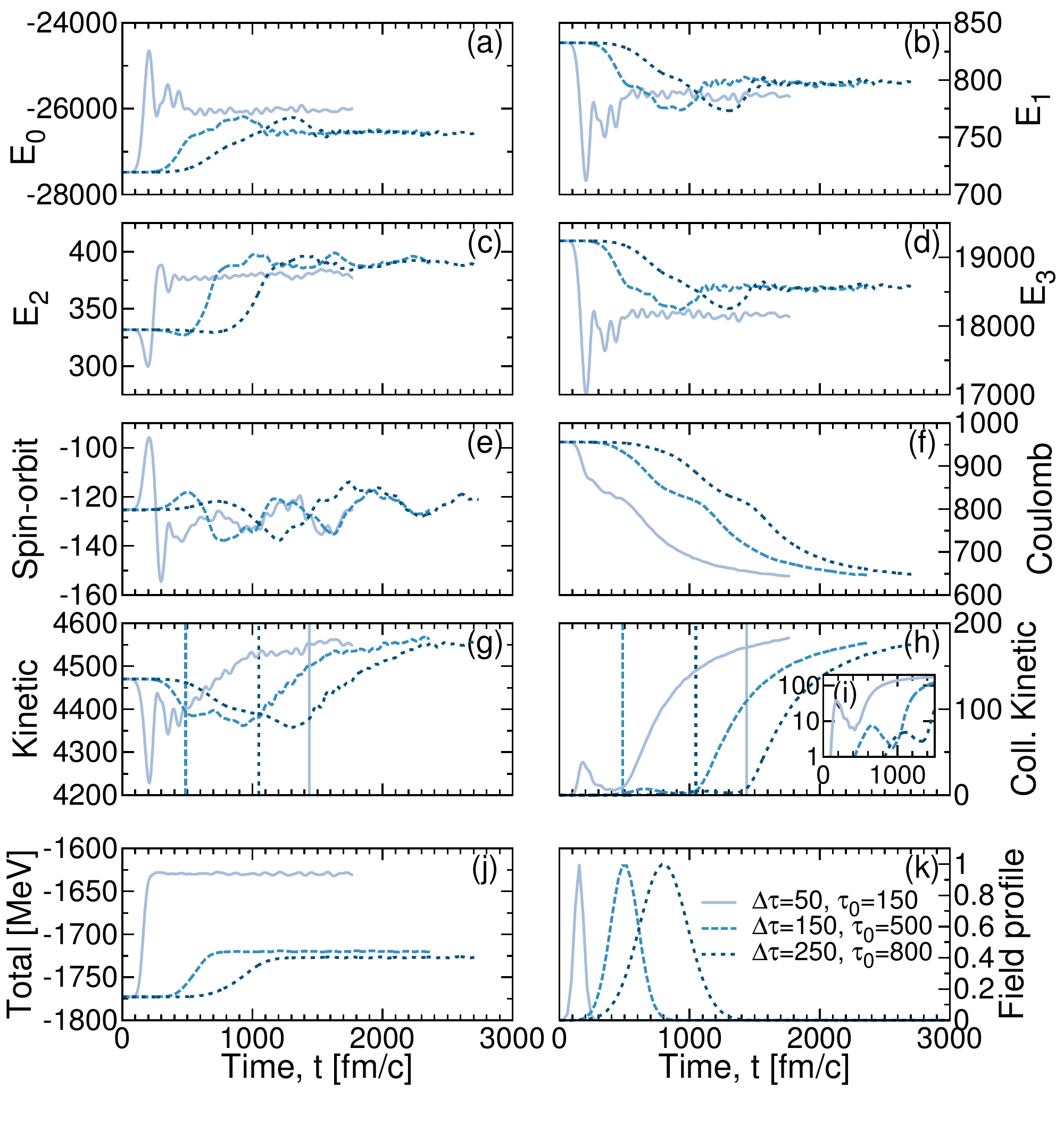} 
\caption{(Color online) (a)-(j): Time evolution of the integrated contributions to the EDF when applying different time-dependent external field to the isomeric state. The (normalized) time profile of the external fields is displayed in panel (k). The boost's strength factor, $\eta$, is adjusted in each case and corresponds to the minimum value needed to induce fission.  Vertical lines in panels (g) and (h) correspond to the point of scission. Units in all panels are MeV. See text and Table~\ref{varied-timeep} for further details.}
\label{different-widths}
\end{center}
\end{figure}  

Time-dependent BIF is sensitive to both the strength of the boost and its time profile. We briefly explore the time profile dependence of BIF in the following. We present the time evolution of the integrated EDF terms for three threshold energy boosts that induce fission when applied to the isomeric state in Fig.~\ref{different-widths}. Case A corresponds to the shortest field (solid line) and is characterized by a width $\Delta \tau=50$ fm/$c$, and a maximum occurring at $\tau_0=150$ fm/$c$. The boost of case B (dashed line) acts for a slightly longer time, $\Delta \tau=150$ fm/$c$, and peaks around $\tau_0=500$ fm/$c$. Finally, the widest case, C (dotted line), has $\Delta \tau=250$ fm/$c$ and $\tau_0=800$ fm/$c$. 

Upon the limit $\Delta \tau\to 0$, instantaneous boosts should be recovered. The EDF for case A behaves similarly to the instantaneous boosts used in the previous section (e.g., Figs.\@ \ref{symm-iso-fiss} and \ref{089-kicks}). The evolution of the nuclear terms in the EDF for case A displays an initial sharp reduction in magnitude of all terms in the functional (other than the collective kinetic energy) at $150$ fm/$c$, corresponding to the centroid of $f(t)$. Beyond this point, oscillations kick in, and the nucleus transitions to a fissioned configuration by $\approx 500$ fm/$c$. The large oscillations in the EDF terms and the large collective kinetic energy suggest a shockwave-type behavior, as in the instantaneous BIF cases. The initial fluctuation in energy upon application of the external field is of a much smaller magnitude than the case of the instantaneous boost, however. Taking the $E_0$ term as an example, the initial spike shows a peak dropping in magnitude by $\approx3000$ MeV, which compares to the $\approx5000$ MeV seen for the corresponding instantaneous boost.

It is interesting to observe that the final values of the $E_0,E_1,$ and $E_3$ terms are approximately equal for the two more gradual profiles (cases B and C). The case with the shortest temporal profile (A) plateaus with a magnitude approximately $1000$ MeV less for both the $E_0$ and the $E_3$ terms and $15$-$20$ MeV less for the $E_1$ term. The differences in the $E_0$ and $E_3$ terms when comparing case A to cases B and C suggest that the final fragments in the latter two cases are less deformed and have a more compact density.

For case A, a sharp drop in the magnitude of the $E_2$ case occurs initially, as was seen when applying instantaneous boosts (Fig.\@ \ref{symm-iso-fiss}). Here, the drop in magnitude is approximately $30$ MeV, which is again much smaller than the drop of approximately $100$ MeV seen for the corresponding instantaneous boost to the isomer. This sharp drop in the $E_2$ term is not seen for cases B and C, which suggests that the process is much more similar to the gradual evolution seen for DIF (no drop in the $E_2$ term was seen in the time evolution for DIF \cite{Goddard2015}). 

The wider temporal profiles of cases B and C show a much more gradual transition in the EDF as the system evolves to fission. No large-amplitude rapid oscillations are seen in the evolution of the EDF terms, and the behavior during the transition to fission is also reminiscent of DIF \cite{Goddard2015}. This suggests that the shockwave-type behavior in the evolution of the densities, which was seen for excitations delivered in a shorter or instantaneous time profile, is not occurring as the system smoothly evolves to fission. 
The initial drops in magnitude in the values of the $E_0, E_1,$ and $E_3$ terms for case C are similar to those of case B. Comparing the initial change in magnitude of the $E_0$ term, a drop of $\approx 1000$ MeV is found for both cases B and C, compared to the drop of $\approx3000$ MeV for case A. In fact, the main difference observed between these two cases and case A is that the evolution of the $E_0,E_1,$ and $E_3$ terms show a recovery in absolute magnitude close to scission.

This exploration demonstrates that the time scale of the energy deposition matter in terms of the resulting fission dynamics. Table \ref{varied-timeep} shows the effect of varying the time-dependent profile $f(t)$ upon the threshold energy required to induce fission. The corresponding threshold energy can also be read off Fig.~\ref{different-widths}(j). We also give the final, extrapolated collective kinetic energies of the fission products, which in this case are two symmetric $^{120}$Ag$_{47}$ fragments. These energies agree within uncertainties: As all three cases produce identical fission fragments, this demonstrates once more that the energy release is dominated by the translational kinetic energy owing to the Coulomb interaction between the fission products. 

\begin{table*}[t!]
\caption{(Color online) Threshold scaling parameters and energies required to induce fission in the isomeric state when applying time-dependent external fields with different temporal profiles. The final collective kinetic energy, corresponding to (mainly) translational kinetic energy has been extrapolated using the technique described in Ref.~\cite{Goddard2015}. }
\centering 
\begin{tabular}{|c|c|c|c|c|c|} 
\hline 
 Case& $\tau_0$      & $ \Delta \tau$        &     Threshold $\eta$        &Threshold Energy  & Extrapolated Coll.                  \\
 &$[$fm/$c$$]$       & $[$fm/$c$$]$      &            &         $[$MeV$]$     &  KE $[$MeV$]$  \\
\hline\hline 
A&150 & 50 &0.0225 $\le \eta_\text{thresh} \le$ 0.0250 &99 $\le E_\text{thresh} \le$ 110 &227(2) \\ \hline
B&500 & 150 & 0.0090 $\le \eta_\text{thresh} \le$ 0.0095 &41 $\le E_\text{thresh} \le$ 52&  221(1) \\ \hline
C&800 & 250 & 0.0070 $\le \eta_\text{thresh} \le$ 0.00725 &33 $\le E_\text{thresh} \le$ 45& 223(3)\\ \hline
\end{tabular}
\label{varied-timeep}
\end{table*}

The application of a time-dependent excitation ensures that the static and dynamic states at $t=0$ are the same, unlike the application of an instantaneous boost, which leads to a mismatch in the initial, excited state. The shortest time profile, case A, requires significantly more energy to induce fission than the more gradual fields of cases B and C. The lowest energy observed to induce fission for case A is $110$ MeV, which is approximately half of that which was required for the instantaneous boost. Longer time profiles demonstrate a significant reduction in the required excitation energy compared to case A, with the lowest energies required found to be $52$ and $45$ MeV for cases B and C, respectively. As the temporal profile of the external field is extended, the energy required to induce fission is reduced. The comparative threshold energies for cases B and C suggest that an adiabatic limit may be approached when using an even more gradual temporal profile for the external field (up until the point of scission).

\subsection{Time-dependent BIF beyond the second barrier peak}
\label{timedep089sec}

\begin{figure}[t!]
\begin{center} 
\includegraphics[width=0.6\linewidth]{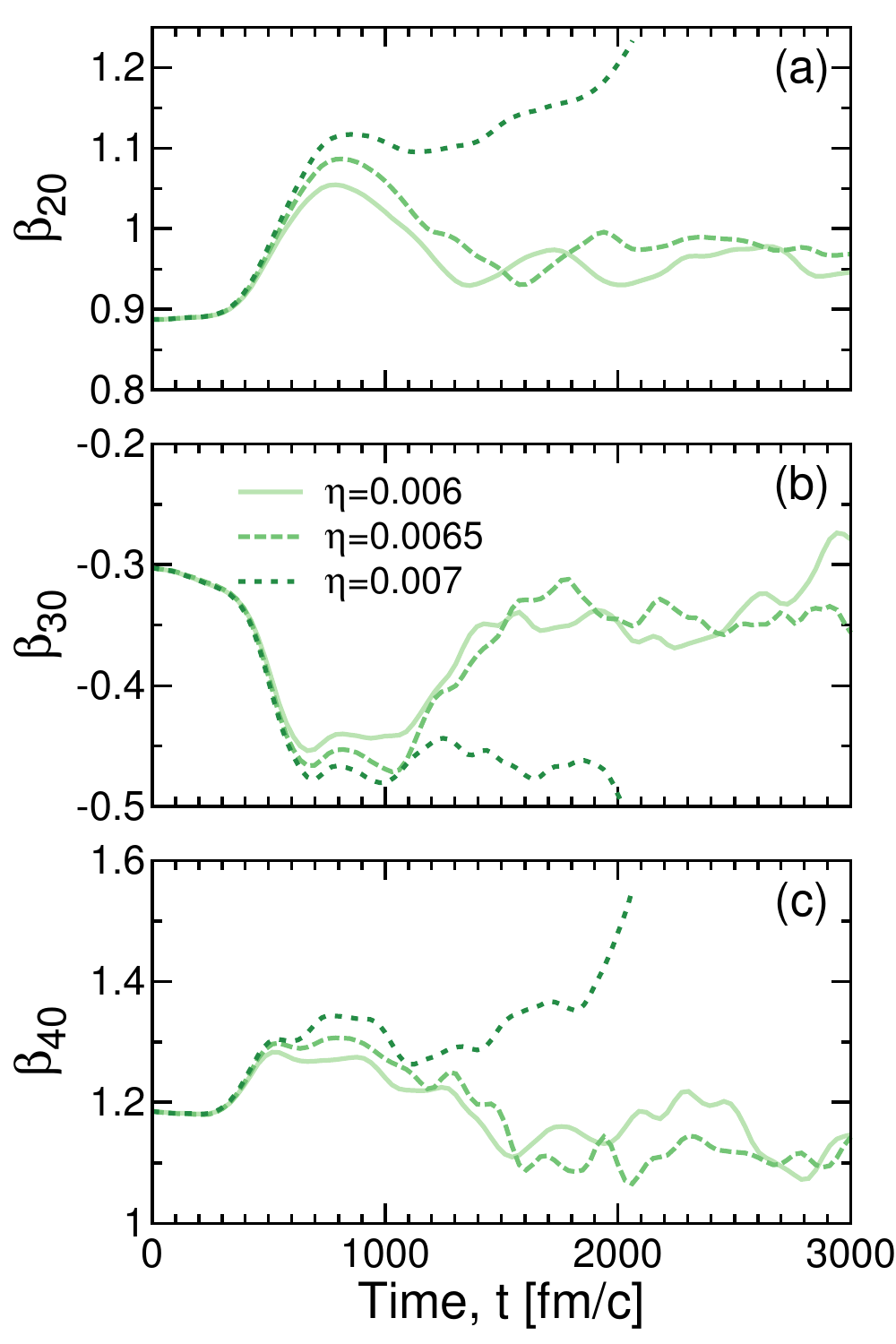} 
\caption{(Color online) Time evolution of (a) quadrupole, (b) octupole, and (c) hexadecapole deformation parameters following time-dependent ($\tau_0=500$ fm/$c$ and $\Delta \tau=150$ fm/$c$) quadrupole excitations upon the initial state with $\beta_{20}=0.890$. The field with scaling parameter $\eta=0.007$ is seen to induce fission, and the evolution of the multipole moments is sharply cut off at the point of scission at $2050$ fm/$c$.}
\label{089-timedep}
\end{center}
\end{figure}

We now briefly discuss the effect of time-dependent boosts on the state with initial deformation $\beta_{20}=0.890$. We use the temporal profile with parameters $\tau_0=500$ fm/$c$ and $\Delta \tau=150$ fm/$c$ as a starting point. Figure \ref{089-timedep} displays the evolution of the multipole moments using various scaling parameters, $\eta$. For $\eta=0.007$, the system is observed to fission, as demonstrated by the rapid increase of elongation in the quadrupole deformation [panel (a)] following the application of the field. The octupole degree of freedom [panel (b)] is also explored owing to the initial mass asymmetry of the static configurations, and for the fissioning case it increases in magnitude from $\beta_{30}= -0.3$ to the region of $-0.5$ at the point of scission.  

\begin{figure*}[t!]
\begin{center} 
\includegraphics[width=0.6\linewidth]{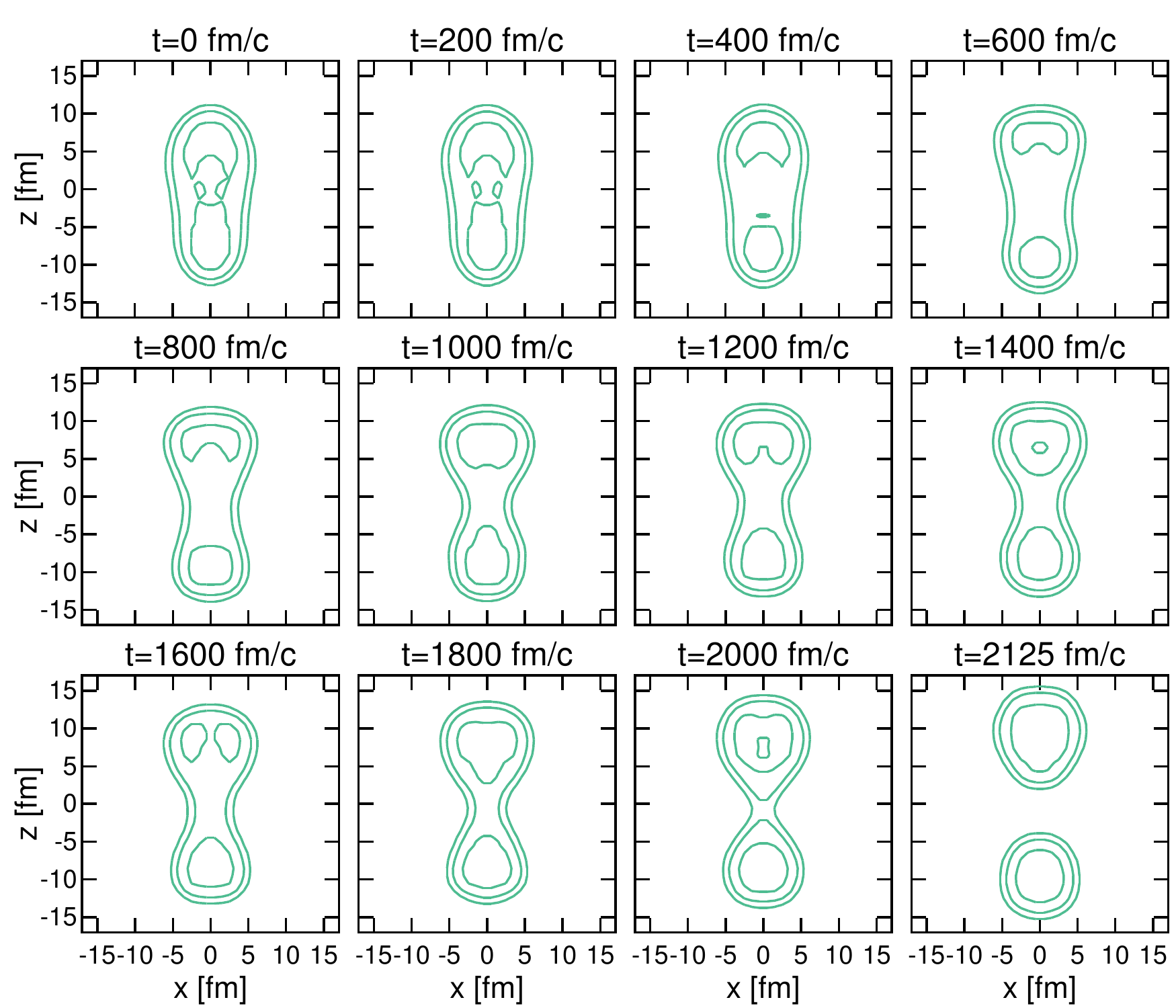} 
\caption{(Color online) Slices of the total particle density for various times, following a time-dependent boost with scaling constant $\eta=0.007$ upon the state with initial deformation $\beta_{20}=0.890$. The isolines are separated by $0.05$ particles/fm$^3$.}
\label{089-timedep-slices}
\end{center}
\end{figure*}

The evolution of the particle density up to the point of scission is displayed in Fig.\@ \ref{089-timedep-slices}. No visible changes are observed in the particle density until $300-400$ fm/$c$, i.e., close to the peak of the boost profile. The scission point lies at $2050$ fm/$c$, a longer time scale than the threshold instantaneous BIF case applied on the same state, which took approximately $1700$ fm/$c$ to fission (see Fig.\@ \ref{089_dens_slices}). The influence of the external field wanes by $800$ fm/$c$. Beyond this point, the neck region smoothly rearranges into the two fission fragments. The relatively small oscillations on the multipole parameters before scission occurs demonstrate that there is a short phase of shape rearrangement before fission. The final fission products are asymmetric. Upon scission, the products are $A_1,Z_1=145.05(5) , 56.32(5)$ and $A_2,Z_2=95.02(5) , 37.69(5)$. To the nearest integer particle number, this gives $^{145}$Ba$_{56}$ and $^{95}$Sr$_{35}$. The mass distributions of this BIF process will be compared to DIF, as well as experimental results, in Sec. \ref{sec:massdistbifdif}. 

\begin{figure}[h!]
\begin{center} 
\includegraphics[width=\linewidth]{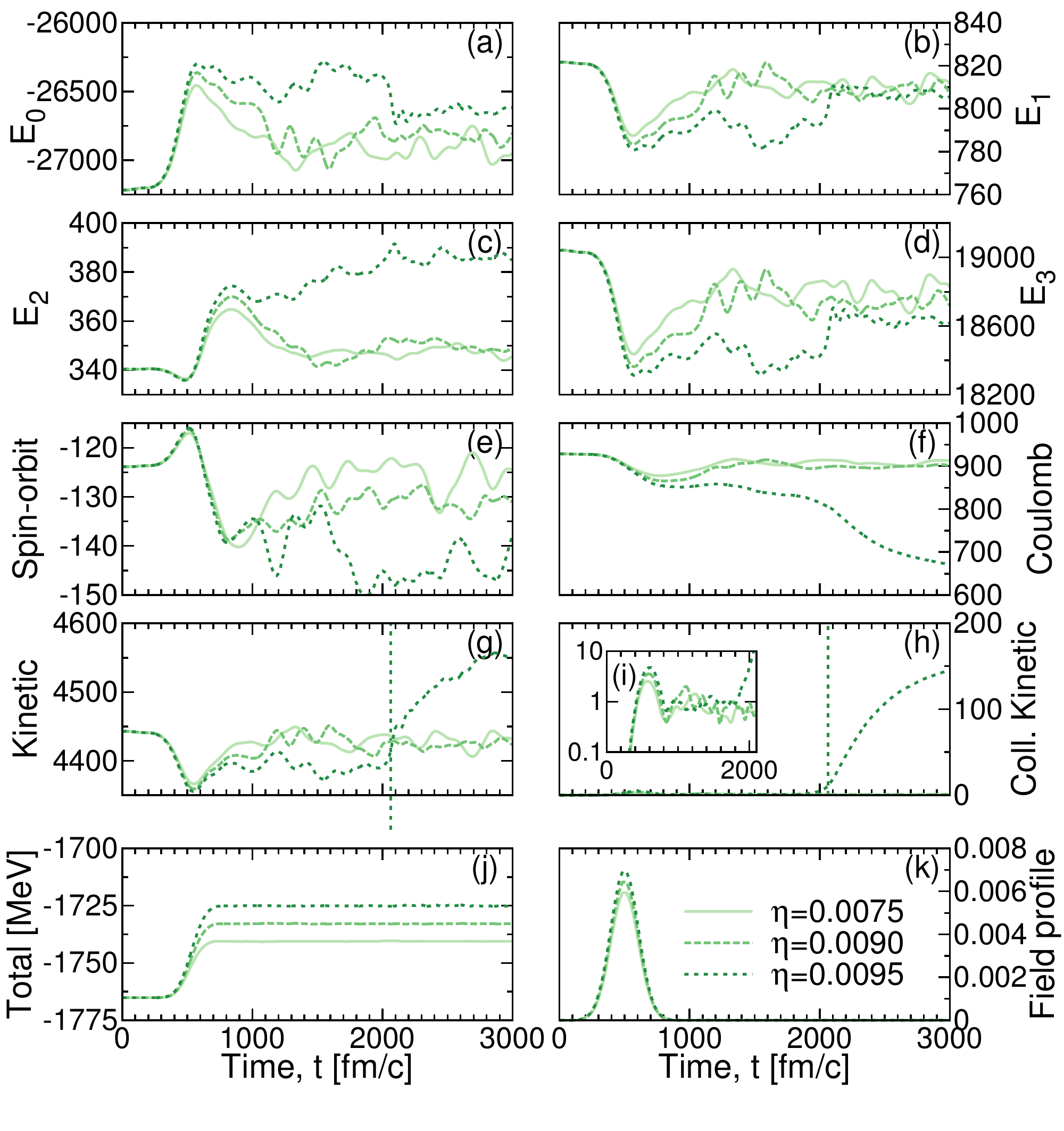} 
\caption{(Color online) (a)-(j): Evolution of the integrated contributions to the EDF following a time-dependent external field boost upon the state with initial deformation $\beta_{20}=0.890$.
The time profile of the external fields is displayed in panel (k). Vertical lines in panels (g) and (h) correspond to the point of scission. Units in all panels are MeV. See text for more details.}
\label{089-timedep-edf}
\end{center}
\end{figure}

We show the decomposed EDF terms for the time-dependent boost upon the state beyond the second barrier in Fig.\@ \ref{089-timedep-edf}. The threshold energy required to induce BIF using the specified time-dependent excitation field is $32\le E_{\text{thresh}}\le40$ MeV. This energy compares to the $225$ MeV required for the minimum energy case of instantaneous BIF for this state (Fig.\@ \ref{0_89_quad}). Once again this demonstrates that, when applying a gradual evolution, a significantly lower threshold energy is required to induce fission. 

The initial drop in magnitude of the $E_0,E_1,$ and $E_3$ terms in panels (a), (b), and (d), respectively, indicate that the nucleus moves to an elongated configuration in the first $500$ fm/$c$. For the fissioning configuration, between $500$ and $2000$ fm/$c$, slight fluctuations are observed in these EDF terms. We take this as a sign that the configuration rearranges owing to the current induced by the excitation. All in all, however, the fluctuations are small. For example, the $E_0$ term varies by less than $250$ MeV while in this elongated configuration. Around the point of scission, which occurs $\approx 2000$ fm/$c$, the characteristic increase in magnitude of the $E_0,E_1$ and $E_3$ terms is observed. The $E_2$ term in panel (c) displays a gradual increase beyond an initial peak at $800$ fm/$c$, as the two fragments form and hence more surface is available. 

As with the other cases where a time-dependent external field has been considered, the collective kinetic energy of Figs.~\ref{089-timedep-edf}(h) and \ref{089-timedep-edf}(i) shows an initial peak near the centroid of the temporal profile of the field. The collective energy subsequently decays as the excitation field ends, and remains constant at around $1$ MeV. This corresponds to the current induced by the field. Its comparatively low value indicates that fewer currents are induced, compared to the instantaneous BIF case. The value is, as expected, closer to that observed for DIF before scission \cite{Goddard2015}. Beyond $2000$ fm/$c$, a rapid increase in collective kinetic energy is found around the scission point as the fragment translational motion sets in. 

As this system takes longer to fission than the other BIF cases, the calculation was terminated at $3000$ fm/$c$, where the fragment separation was only $75$ fm. We can find the corresponding collective energy using the same method as in Ref.~\cite{Goddard2015}, and find a translational energy at large times of $207(9)$ MeV. 

\section{DIF and BIF fission fragment masses}
\label{sec:massdistbifdif}

In this section, we summarize the results regarding the masses of the final fission fragments that have been obtained within our dynamical simulations. We compare these to the experimental neutron-induced data of Ref.~\cite{Jfis}. We note, however, that this comparison is only indicative, as our theoretical simulations are limited by a variety of factors. First, we only include BCS pairing in the initial state, then keep the occupations fixed and hence do not consider explicit superfluid dynamics \cite{Ste11,Scamps2015,Tanimura2015}. Second, beyond-mean-field correlations can play an important role on both the shape of the PES and the fission dynamics \cite{Bernard2011}. Third, we take our mass numbers as the nearest integer to the actual (non-integer) particle numbers for the fission fragments. In other words, we do not project our final fragments into a good particle number \cite{Sim10}, and hence do not have access to a mass distribution \cite{Scamps2015}.

In spite of these limitations, the TDHF fission trajectories are indicative of an average fragment behavior. More interestingly, as opposed to static calculations, the BIF fragments produce a \emph{variety} of final fragments depending on the initial state in the PES and the strength and time modulation of the boost. For BIF, we only consider cases where the static configuration had mass asymmetry. Owing to the symmetric nature of the excitation fields applied, an initial configuration with no octupole deformation is unable to explore this degree of freedom and necessarily leads to symmetric fission fragments.

\begin{figure}[h!]
\begin{center} 
\includegraphics[width=0.6\linewidth]{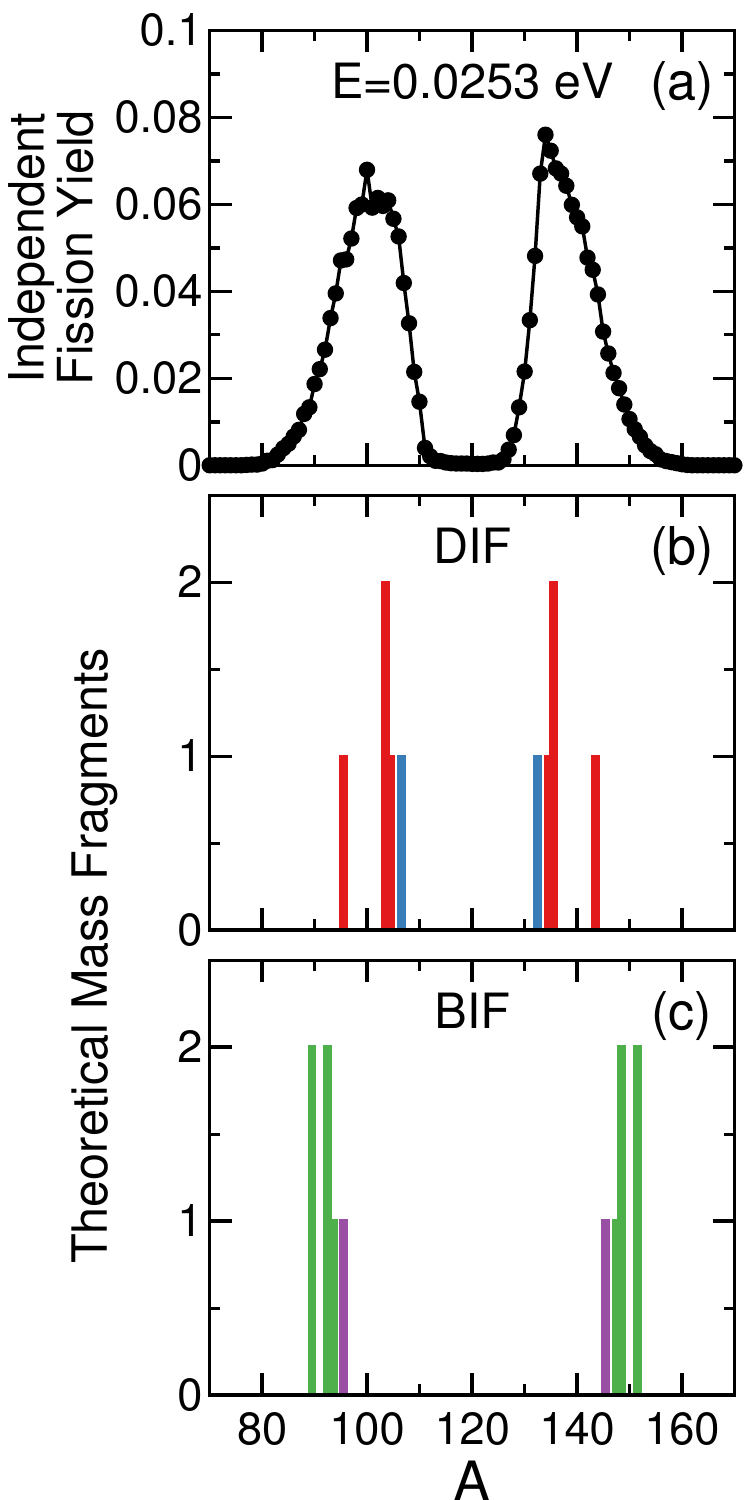} 
\caption{(Color online) (a) Experimental independent fission yields for neutron-induced fission at $E=0.0253$-eV energies. The data are from Ref.~\cite{Jfis}. (b) Theoretical mass fragments obtained from the DIF process in Ref.~\cite{Goddard2015}. The red bars correspond to the binned TDHF results and the blue bar corresponds to the static two-fragment mass split. (c) Theoretical mass fragments obtained from different BIF processes upon the state with initial deformation $\beta_{20}=0.890$.}
\label{kick-mass-dist}
\end{center}
\end{figure} 

Figure \ref{kick-mass-dist} displays a comparison between the experimental data obtained from low energy neutron-induced fission in panel (a) and theoretical dynamical calculations. Whereas panel (b) displays the masses obtained from the DIF cases presented in Ref.~\cite{Goddard2015}, the BIF cases are examined in panel (c). The green bars correspond to the masses obtained using instantaneous excitations (Table \ref{089_boosts}), and the purple bars correspond to the sample time-dependent excitation field of the previous section.

DIF fission products fall well within the experimentally obtained mass distributions. We note, in particular, that DIF produces more asymmetric mass fragments than the corresponding adiabatic two-fragment pathway (red bars). In contrast, when applying instantaneous boosts to the static state with deformation $\beta_{20}=0.890$, the resulting masses fall on the edge of the experimentally obtained results. In other words, BIF produces even more asymmetric fragments than DIF does. Instantaneous boosts, as opposed to time-dependent boosts, yield the most asymmetric fragments. These correspond to the green bars in panel (c). We note, however, that there is no clear pattern on the energy dependence of the mass fragments. For instance, BIF with the lower threshold energy of $E=225$ MeV produces the same fission fragments as a higher-energy $E=400$-MeV boost. 

As we have seen so far, the time-dependent BIF process is somewhat related to DIF for wide enough boost profiles. The purple bar in panel (c) shows the fragment masses for the  time-dependent BIF event with $\Delta\tau=150$ fm/$c$ and an energy deposition of $\approx 40$ MeV. The corresponding masses lie closer to the peak of the experimental distribution than the other BIF examples. In time-dependent BIF, the collective energy deposited into the system is rapidly transferred into the nuclear terms of the EDF. As a consequence, the overall collective energy is relatively small until the point of scission. In a sense, the pre-fission system can reconfigure and absorb the energy more ``adiabatically" than the instantaneously boosted system does. It is thus not surprising that the corresponding fission fragments lie closer to the DIF results. Moreover, the comparison of instantaneous BIF to temporally extended BIF demonstrates that the time scale for the energy deposition has important consequences regarding the fission dynamics.

The time scale is not the only parameter that determines the final state. The strength of the boost and the corresponding deposited energy also determine the fate of the fissioned state. As we have already discussed, different energies delivered in an instantaneous boost will lead to different fission fragments. The main reason underlying this non-linearity is the fact that the energy of the boost is quickly absorbed into the nuclear part of the functional. As a consequence, the nucleus rapidly changes shapes and oscillates violently as soon as the boost is imparted. If octupole degrees of freedom are explored, for instance, the system can reconfigure dynamically into different excited configurations if it has been excited at different energies. In turn, these configurations will produce different fission fragments. However, because the fission mechanism is induced by the shape oscillations and the onset and dominance of Coulomb repulsion, but not by the boost itself, the memory of the initial boost properties is lost in the dynamics. The final fission fragments are thus relatively similar, in spite of the different quadrupole boosts that started the dynamics. Moreover, because most of the collective energy of the system is transferred into the kinetic energy of the fragments via the Coulomb repulsion, one also finds that the final fission fragment energies are very similar. 

It is therefore not so surprising that the threshold energy of $225$ MeV required for an instantaneous BIF event from the state with $\beta_{20}=0.890$ is at least an order of magnitude larger than that required to induce fission experimentally. For example, photo-fission may be induced in $^{240}$Pu using a $12$-MeV end-point energy bremsstrahlung source \cite{Thi81} (although such a process corresponds to a dipole excitation, rather than the quadrupole excitations examined here). The threshold energy scale is also larger than the fission barriers of $^{240}$Pu. In a physical scenario, particularly if slow neutrons are involved, the energy delivery process is unlikely to be instantaneous. We have already seen that time-dependent BIF requires far less energy to produce a final fissioned state, particularly if the energy delivery process is slow (see Table~\ref{varied-timeep}).

It is a well-known fact that the neutron-induced fission fragment distribution of Pu becomes more symmetric as the energy of the neutron increases \cite{Byrne1994,Jfis}. In other words, the relative intensity of symmetric fission increases with neutron energy. Naively, one would associate more energetic neutrons to faster energy deposition, which could be akin to instantaneous BIF simulations. In contrast, we find that instantaneous BIF induces more asymmetric products than its time-dependent counterpart, or than the DIF process, for that matter. Fast neutrons, however, are unlikely to be well represented by a quadrupole boost. The geometry of the energy-deposition mechanism is different. In particular, even a very fast neutron will hit the surface of the nucleus first, whereas the boosts that we have implemented affect the nucleus throughout its density (see Fig.~\ref{is_quad_curr}). In a TDHF picture, one could presumably simulate neutrons in terms of moving wave packets or, in a picture closer to what we have developed, as external density fields \cite{Assie2009}. Further research in this direction would produce a more microscopic insight into neutron-induced fission phenomena.

\section{Conclusions}
\label{sec:conclusions}

We have presented an analysis of induced fission reactions using TDHF, implemented in {\sc sky3d}. We have focused our study on the benchmark fission isotope $^{240}$Pu and have investigated two mechanisms to excite the nucleus energetically and push it towards a scission phase. We have explored fission from two physically different states. One is the fission isomer at $\beta_{20}=0.682$, in between the two fission barriers. The other is a state that lies just beyond the second barrier peak, with static deformation $\beta_{20}=0.890$, but before the intersection of the one- and the two-fragment pathways. The isomeric state lies in the region that we have dubbed ``forbidden" for spontaneous fission. The second state, in contrast, lies in an inhibited area, where DIF does not occur. In spite of the physical differences underlying the two initial states, one conclusion of our implementation of the BIF process is that very similar fission processes occur for both states.

Further, we have implemented two different types of quadrupole shaped boosts. Owing to the gauge invariance of the Skyrme EDF, instantaneous boosts deliver all their strength in the form of kinetic collective energy. The amount of energy that is delivered can be adjusted  at will (see the Appendix). We find that minimum threshold excitation energies of the order of $\approx 200$ MeV are required to induce fission. This is a significantly greater energy than the corresponding static fission barrier heights $\approx10$ MeV. In the BIF process, the excitation energy is first deposited as collective energy and then absorbed within the first $100-200$ fm/$c$ into the nuclear potential terms of the EDF. A significant current is induced, and a violent evolution of the state ensues. The nucleus oscillates significantly in shape and finds a dynamical doorway towards a fissioned state. This is in contrast to the gradual rearrangement of the densities observed for the DIF process.

The second type of boost that we have implemented are modulated in time. We have applied a quadrupole excitation field with a Gaussian time profile of varying widths. For a profile with a width of $150$ fm/$c$, the collective kinetic energy is delivered mildly. The nucleus thus readjusts to the quadrupole pull by transferring energy into the nuclear EDF. As a consequence, the pathway to fission is relatively smoother. The energy required to induce fission, for instance, is significantly reduced. When exciting the fission isomer, about $45$ MeV is enough to generate fission with a time-dependent energy deposition of width $250$ fm/$c$. One finds several similarities between the BIF and the DIF processes. Both show a smooth, gradual evolution of the densities up to around the point of scission. There is a small amount of currents, evidenced by a small collective energy, that eventually reconfigure the nucleus and lead it to fission. When evolving the state using temporally extended BIF, the time scale for scission is approximately $2000$ fm/$c$, during which the rearrangement of the energies and densities is far less extreme than the corresponding instantaneous BIF processes. This time scale compares well to the DIF case with the least deformed initial configuration. 

The time scales for fission, as well as the final fragment masses, depend on the strength of the boost and the time dependence of the energy-deposition scheme. Instantaneous boosts with different energies, for instance, can produce the same final fragments. When exciting the state at  $\beta_{20}=0.890$, a boost with $225$ MeV produces the same final fragments as a boost with $400$ MeV, but fissions almost $1000$ fm/$c$ later. In part, this is attributable to the fact that the violent shape oscillations induced by the very energetic initial excitations are qualitatively similar. The nucleus is in a dynamically excited state, and in a violent process of exploration of different shapes, until it eventually finds a favorable fission path. 

In our approach, mass asymmetric fragments are only reached from states that are initially octupole deformed. The fission fragment masses obtained with instantaneous BIF methods are somewhat more asymmetric than their DIF counterparts, and lie further away than the corresponding experimental peaks.  Time-dependent boosts produce results that lie in between the two. Again, these are a reflection of the different dynamics explored with the two BIF methods. All in all, these results illustrate the richness and variety of final states that can be accessed by means of a dynamical simulation of fission, in contrast to the adiabatic picture. 

A variety of extensions of this work could prove of use in future studies of fission within microscopic  time-dependent approaches. From a theoretical perspective, the inclusion of dynamical superfluid effects would certainly provide more realistic simulations. Similarly, extensions beyond traditional TDHF calculations, via projection methods, could provide access to mass distributions and widths. At a more practical level, one could investigate boosts with varying geometries, which would presumably provide a more comprehensive set of fission fragment masses, as well as indicating which are the most appropriate collective coordinates with which to induce fission. To simulate more realistic induced fission events, one could also investigate moving energy-deposition sources, instead of spatially static boosts. The use of wave packets or external fields to simulate, for instance, slow neutron absorption could also provide a more detailed insight into induced fission mechanisms. Finally, in keeping with the observation of a ternary fission event for instantaneous BIF, it would be interesting to study the formation process and dynamics of multifragment fission events.

\begin{acknowledgments}
This work is supported by STFC, through Grants No. ST/I005528/1, No. ST/J000051/1 and No. ST/J500768/1. This research made use of the STFC DiRAC HPC cluster.  
\end{acknowledgments}

\appendix
\section*{Appendix: Boost energy}
\label{app1}

The adopted EDF of the nuclear system is given by \cite{Mar13}
\begin{equation}
E ={E_\text{Kin}}+E_\text{Skyrme} +E_\text{Coulomb} \, .
\end{equation}
An instantaneous velocity boost applied by the transformation $\varphi(\bm r) \to e^{i \eta \phi(\bm r)} \varphi(\bm r)$ is a local gauge transformation. The Skyrme and the Coulomb terms of the EDF are locally gauge invariant \cite{Dob96,Dob05}. Consequently, an instantaneous boost only leads to an increase in kinetic energy,
\begin{equation}
\Delta {E}_\text{kin} = \frac{\hbar^2}{2m} A \eta^2 \langle \,  |\nabla \phi(\bm r)|^2 \, \rangle \, ,
\end{equation}
where $A$ is the integrated particle density and the average, $\langle \cdot \rangle$, is taken over the nuclear density profile. $\Delta E_{\text{kin}}$, which we dub excitation energy, is thus only delivered as collective kinetic energy, and $\eta \nabla \phi(\bm r)$ can be interpreted as a velocity field \cite{Dob05}. 
By rearranging for $\eta$,
\begin{equation}
\eta = \sqrt{\frac{\Delta {E}_\text{kin}}{\frac{\hbar^2}{2m}A \langle \, |\nabla \phi(\bm r)|^2\, \rangle}} \,,
\end{equation}
the magnitude of the boost may be fixed for a given $\Delta E_{\text{kin}}$.

As an example, to excite the nucleus by adding collective kinetic energy via a quadrupole velocity field, one may consider
\begin{equation}
\phi(\bm r) = \eta(2z^2 - x^2 - y^2) \, .
\end{equation}
Rearranging for $\eta$, one finds:
\begin{equation}
\label{etainst}
\eta = \frac{1}{2}\sqrt{\frac{\Delta {E}_\text{kin}}{\frac{\hbar^2}{2m}A \langle \, 4x^2 + 4y^2 + 16z^2 \, \rangle}} \, ,
\end{equation}
which may be used to add a precise amount of excitation energy, $\Delta {E}_\text{kin}$, to an initial state with known density profile.

The effect of a time-dependent boost is more difficult to compute. As soon as some collective kinetic energy is introduced into the system, dynamics will transfer energy into other degrees of freedom, e.g., shape degrees of freedom. 
A change in shape will thus impact the nuclear terms of the EDF, and it becomes difficult to find a closed form for the change of the different terms over time. We note, however, that the boost can also be thought of as an external field acting over time; cf., Eqs.~(\ref{eq:hprime}) and (\ref{eq:uprime}). Within TDHF, this external field will produce a change in the (otherwise conserved) total energy. One can thus read off the total amount of deposited energy by comparing the initial and the final (e.g., after the boost is exhausted) total energy of the system.

\bibliographystyle{apsrev4-1}
\bibliography{biblio}

\end{document}